\RequirePackage{ifpdf}

\documentclass[letterpaper]{JHEP3}
\usepackage{amsmath, amssymb, mathtools, stmaryrd}
\usepackage{amsthm}
\usepackage{amsfonts}
\usepackage{epsfig,multicol,float}

\def\ba{\begin{align}}
\def\ea{\end{align}}
\def\be{\begin{equation}}
\def\ee{\end{equation}}
\def\nn{\nonumber}
\def\bea{\begin{eqnarray}}
\def\eea{\end{eqnarray}}

\def\exd{{\rm d}}

\def\pref#1{(\ref{#1})}

\newcommand{\roughly}[1]{\mathrel{\raise.3ex\hbox{$#1$\kern-0.85em
  \lower1ex\hbox{$\sim$}}}}

\newcommand{\gsim}{\roughly>}

\def\Fd{{\widetilde F}}
\def\Gd{{\widetilde G}}

\def\cA{{\cal A}}
\def\cD{{\cal D}}
\def\cE{{\cal E}}
\def\cF{{\cal F}}
\def\cG{{\cal G}}
\def\cJ{{\cal J}}
\def\cO{{\cal O}}
\def\cT{{\cal T}}

\def\ssB{{\scriptscriptstyle B}}
\def\ssF{{\scriptscriptstyle F}}
\def\ssV{{\scriptscriptstyle V}}

\makeatletter
\def\x@arrow{\DOTSB\Relbar}
\def\xlongequalsignfill@{\arrowfill@\x@arrow\Relbar\x@arrow}
\newcommand{\xlongequal}[2]{%
	\ext@arrow 0099\xlongequalsignfill@{#1}{#2}}
\makeatother

\title{AdS/QHE: Towards a Holographic Description of Quantum Hall Experiments}

\author{Allan Bayntun,${}^1$ C.P.~Burgess,${}^{1,2}$ Brian P. Dolan${}^{3,4}$ and Sung-Sik Lee${}^{1,2}$\\

${}^1$Department of Physics \& Astronomy, McMaster University\\ \qquad 1280 Main Street West, Hamilton ON, Canada.\\

${}^2$Perimeter Institute for Theoretical Physics\\
\qquad 31 Caroline Street North, Waterloo ON, Canada.\\

${}^3$Dept. of Mathematical Physics, National University of Ireland, Maynooth, Ireland.\\

${}^4$School of Theoretical Physics, Dublin Institute for Advanced Studies\\
\qquad 10 Burlington Rd., Dublin, Ireland.\\

}

\abstract{Transitions among quantum Hall plateaux share a suite of remarkable experimental features, such as semi-circle laws and duality relations, whose accuracy and robustness are difficult to explain directly in terms of the detailed dynamics of the microscopic electrons. They would naturally follow if the low-energy transport properties were governed by an emergent discrete duality group relating the different plateaux, but no explicit examples of interacting systems having such a group are known. Recent progress using the AdS/CFT correspondence has identified examples with similar duality groups, but without the DC ohmic conductivity characteristic of quantum Hall experiments. We use this to propose a simple holographic model for low-energy quantum Hall systems, with a nonzero DC conductivity that automatically exhibits all of the observed consequences of duality, including the existence of the plateaux and the semi-circle transitions between them. The model can be regarded as a strongly coupled analog 
of the old `composite boson' picture of quantum Hall systems. Non-universal features of the model can be used to test whether it describes actual materials, and we comment on some of these in our proposed model. }

\begin{document}

\section{Introduction}

Applications of AdS/CFT duality \cite{AdSCFT, AdSCFTrev1, AdSCFTrevs} to condensed matter physics \cite{AdSCMTrevs} carry a whiff of a fishing expedition. The goal is to explore the properties of strongly interacting conformal field theories (CFTs) using their calculable gravity duals in anti-de Sitter space (AdS). The jackpot would be to find a model that describes a strongly correlated system of real electrons; systems that have resisted approaches using other theoretical tools. Without a systematic way to derive the magic CFT directly from underlying electron dynamics one throws theoretical darts into field space, hoping to find that right `hyperbolic cow.'

Like any fishing expedition, it always helps to have some local guidance towards the good fishing holes. What would be useful are a set of simple properties, like symmetries, that are known to be prerequisites for a successful description of a particular system. Knowledge of these properties could help guide the search for theories that are relevant to life in the lab.

In this paper we argue that quantum Hall systems \cite{GirvinReview} are likely to be profitable places to fish, for two reasons. First, they involve strongly correlated electrons, and for decades have been a source of new experimental phenomena requiring theoretical explanation. But their phenomenology also points to symmetry properties that seem relatively easy to find in an AdS framework, and these symmetries can help narrow down the search for the killer model. Our purpose is threefold: to briefly summarize the relevant phenomenology and the symmetries to which we believe they point; to propose a particular class of AdS/CFT models that captures these symmetries; and to identify a class of tests for such models that go beyond the implications of the symmetries, to be used to home in on an experimentally successful model.

The symmetries of interest are not symmetries in the usual sense. Rather they are a large group of duality transformations that appear to map the various quantum Hall states into one another, and which commute with the RG flow of these systems at very low temperatures as one approaches the many quantum Hall plateaux. In particular, we summarize in \S2\ the evidence for the existence of discrete duality transformations of this type, acting on the ohmic ($\sigma_{xx}$) and Hall ($\sigma_{xy}$) conductivities according to the rule
\be \label{eq1}
 \sigma := \sigma_{xy} + i \sigma_{xx}
 \to \frac{ a \, \sigma + b}{ c \, \sigma + d} \,,
\ee
where $a$, $b$, $c$ and $d$ are integers satisfying the $SL(2,Z)$ condition $ad-bc = 1$, but with $c$ restricted to be even. The consequences of this symmetry include a number of well-measured effects for quantum Hall systems, including the kinds of fractional states that can arise as attractors in the low-energy limit; which states can be obtained from which others by varying magnetic fields; detailed predictions for some of the trajectories through the conductivity plane as the temperature, $T$, and magnetic field, $B$, are varied; as well as others.

\S2\ describes the qualitative picture: at low energies the flow in coupling-constant space appears to be onto a two-dimensional surface that governs the final approach to the various quantum Hall ground states. The flow in this two-dimensional surface is constrained by the emergent symmetry, eq.~\pref{eq1}, and can be traced experimentally by varying both $B$ and $T$. What is missing is a simple class of candidate models to describe this two-dimensional flow, including the emergent duality. Besides providing an existence proof, having such a model in hand would allow this picture to be sharpened considerably by allowing its implications to be explored in more detail.

What is encouraging is that there is good evidence that transformations like eq.~\pref{eq1} arise quite generically in CFTs having conserved currents in two spatial dimensions \cite{PVD, Witten}. Furthermore, the development of the AdS/CFT correspondence has opened up new tools for exploring strongly interacting 2+1 dimensional CFTs, with the conserved current being dual on the gravity side to an electromagnetic gauge potential. In this language the dual version of the CFT's discrete dualities are rooted in electric-magnetic duality. Applications of these tools to condensed matter remain very promising \cite{AdSCMTrevs}, and studies of the simplest holographic charge-carrying systems do reveal a number of duality-related features \cite{HKSS, holoSCdual}.

The most striking examples to emerge to date of explicit systems with symmetries like eq.~\pref{eq1} are those based on dilatonic black branes \cite{dilaton1,dilaton2} --- briefly described in \S3 --- for which the electric-magnetic duality is also accompanied by an action on the dilaton and axion fields (as in Type IIB supergravity in 10 dimensions). If the duality symmetries provide a good guide, it is among this type of AdS/CFT system that a description of low-energy quantum Hall systems is likely to reside. (See also \cite{holographicHall} for other discussions of quantum Hall systems within an AdS/CFT context.) The main drawback of the simplest dilaton black brane models is their prediction of vanishing DC ohmic conductivity at nonzero temperature. This clearly cannot describe real quantum Hall systems, for which the evidence for eq.~\pref{eq1} relies almost exclusively on DC charge-transport properties.

For this reason we propose, in \S4, a slight modification of this model, following a recent proposal \cite{StrangeM} for strange metal holography. In this proposal the field content of the AdS dual is the same as for ref.~\cite{dilaton2} --- {\em i.e.} gravity, Maxwell field, dilaton and axion --- but with the Maxwell kinetic term described by the (dilaton) Dirac-Born-Infeld (DBI) action rather than the dilaton-Maxwell action. The DBI action shares the desired duality of the dilaton-Maxwell action, but also allows nonzero DC conductivities with which to probe its implications. Following \cite{StrangeM} we treat the charge carriers in the probe-brane approximation, coupled to a black brane that we treat as two separate charged and uncharged cases. (The brane geometry can also be chosen to have Lifshitz form if it is desired to introduce different powers, $z$, for temporal and spatial scalings.) Physically, this corresponds to regarding the charge carriers as perturbations to the CFT described by the black 
hole.

Finally, \S5\ describes a number of the model's predictions that go beyond its basic duality properties. These are tests whose comparison with experiment ultimately provide the scorecard of how successful this, or any other, model is. In particular this section identifies the parameters that control the scaling exponents that are measured in transitions between Hall plateaux and between plateaux and the Hall insulator (see \S2\ for details). Yet the most important message is probably not whether this model succeeds or fails; rather what is important is that there is now a good class of AdS/CFT models having duality properties that closely resemble those of real quantum Hall systems. Hopefully the fishing will be good.

\FIGURE[ht]{ \includegraphics[scale=0.4]{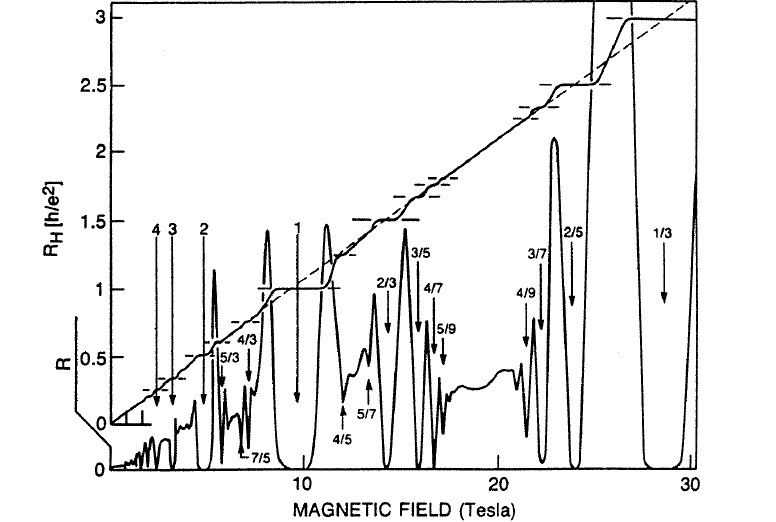}
\caption{Experimental traces of the Hall and ohmic resistances for a quantum Hall system, reproduced from ref.~\cite{Stormer}.
} \label{Fig:plateaux} }

\section{Quantum Hall systems}

This section has a two-fold purpose. First, it is meant to summarize briefly the experimental evidence for duality in quantum Hall systems, since this motivates using duality to guide the search for theoretical descriptions. This is followed by a description of the low-energy effective theory, including a discussion of the `composite boson' model that allows some intuition for the potential origin of the underlying duality transformations, and are the precursors for the effective theories described in the remainder of the paper.

\subsection{Evidence for duality}

Quantum Hall systems are remarkable in a number of ways, not least of which is the very existence, stability and precision of the various plateaux --- see Fig.~\ref{Fig:plateaux} --- for which the ohmic DC conductivity, $\sigma_{xx}$, vanishes\footnote{Notice that the vanishing of the conductivity, $\sigma_{xx}$, also ensures the same for the resistivity, $\rho_{xx}$, when the Hall conductivity is nonzero, $\sigma_{xy} \ne 0$.} and the DC Hall conductivity, $\sigma_{xy}$, is quantized (in units of $e^2/h$, or $e^2/2\pi$ when $\hbar = 1$). The quantized value for $\sigma_{xy}$ at a plateau is always consistent with a fraction, $p/q$, and (with a very few exceptions, to do with other kinds of physics) $q$ is odd.

\FIGURE[ht]{ \includegraphics[scale=0.6]{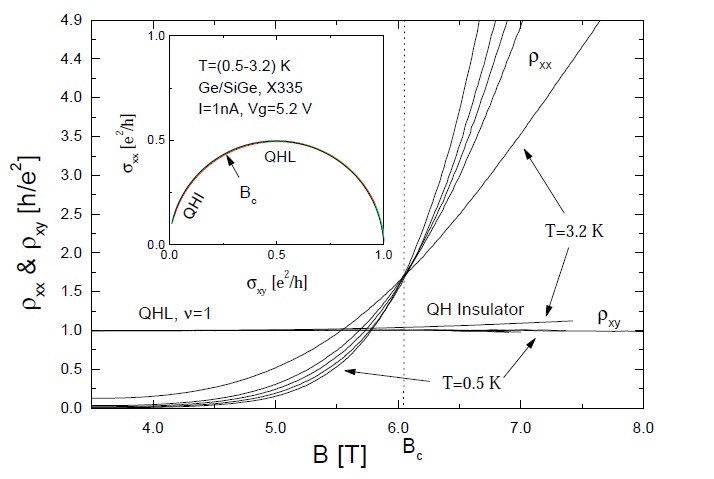}
\caption{Evidence for the semi-circle law in the trace of the conductivities during a transition between two plateaux, reproduced from ref.~\cite{Hilke}.
} \label{Fig:semicircle} }

\subsubsection*{Some relevant experiments}

The evidence for duality lies in the nature of the transitions that are observed to occur between these plateaux as $B$ is changed, as well as in the details of how they are approached at low temperatures. For example:

\vspace{6mm}\noindent
{\em Selection Rule:} As Fig.~\ref{Fig:plateaux} shows, for clean samples a large number of plateaux can be accessed with changing magnetic field, but there is a pattern to the plateaux that are found adjacent to one another. Whenever two plateaux, labeled by the fractions $p/q$ and $r/s$ are clearly adjacent, they satisfy $|ps - qr| = 1$. There are only two exceptions to this rule in Fig.~\ref{Fig:plateaux} --- $\frac53 \to \frac75$ and $\frac45 \to \frac57$ --- but in both cases these two plateaux are not cleanly adjacent to one another.

\vspace{4mm}\noindent
{\em Semi-circle Law:} The precise shape of the resistance curves between two well-defined adjacent plateaux becomes striking once it is drawn as a curve in the $\sigma_{xx}-\sigma_{xy}$ plane. A sample experimental trace of this appears in the inset of Fig.~\ref{Fig:semicircle}, which shows that the trajectory sweeps out a precise semi-circle, with centre midway between the two plateaux.

\FIGURE[hb]{ \includegraphics[scale=0.6]{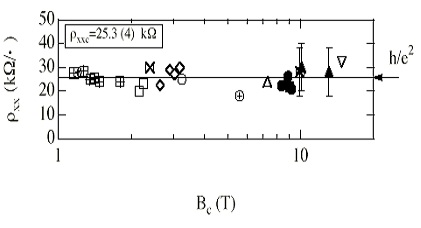}
\caption{Evidence for universality of critical resistivity, $\rho_{\star\,xx} = \rho_{xx}(B_c)$, from ref.~\cite{Yang}.
} \label{Fig:rhocrit} }

\vspace{4mm}\noindent
{\em Critical points:} The remainder of Fig.~\ref{Fig:semicircle} shows the dependence of the resistivities on magnetic field, for several choices of temperature. These show that at fixed $B$, the resistivity $\rho_{xx}$ (and so also, for nonzero $B$, $\sigma_{xx}$) fall to zero with decreasing temperature near a plateau. But for very large magnetic fields, eventually the ohmic resistivity {\em grows} as the temperature falls, defining a regime called the {\em quantum Hall insulator} \cite{Insulator}. The crossover between these two regimes defines a critical magnetic field, $B_c$, for which $\rho_{xx}$ is temperature-independent (also visible in Fig.~\ref{Fig:semicircle}). The value, $\rho_{\star\,xx} = \rho_{xx}(B_c)$, of the resistivity at the critical field appears to be universal inasmuch as it is largely sample-independent. For the transition from the $\sigma_{xy} = 1$ state to the Hall insulator the critical resistivity takes on a value consistent with $\rho_{\star\,xx} = h/e^2$ --- see Fig.~\ref{Fig:
rhocrit}. (As both Figs.~\ref{Fig:semicircle} and \ref{Fig:rhocrit} show, the universality of this critical value is not completely clear in all experiments. The interpretation of this is examined more carefully in \cite{Gamma2}, where it is found that this implication of duality symmetries can be more sensitive to perturbations (like Landau-level mixing) than are some of the others (like the semicircle law).)

\FIGURE[ht]{ \includegraphics[scale=0.5]{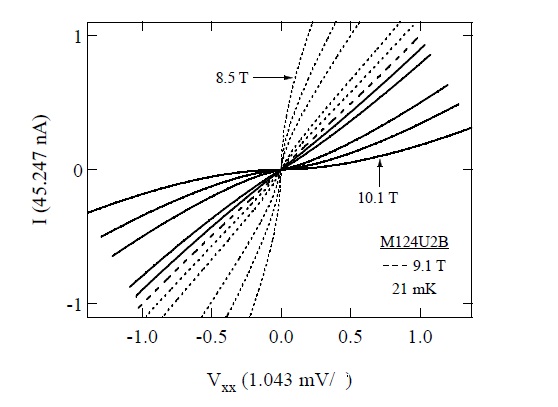}
\includegraphics[scale=0.5]{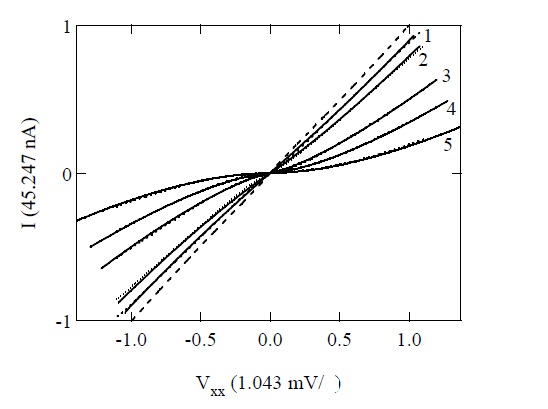}
\caption{Evidence for the duality, $\rho_{xx} \to 1/ \rho_{xx}$, for resistivities equally spaced (in units of filling fraction, $\Delta \nu$) from the critical field, reproduced from ref.~\cite{Nonlin}.
} \label{Fig:dual} }

\vspace{4mm}\noindent
{\em Duality:} The dependence on temperature and magnetic field of $\rho_{xx}$ in a transition from a plateau to the Hall insulator is measured to be consistent with
\be \label{phenodualform}
 \rho_{xx} = \rho_{\star\,xx} \, \exp \left[ - \frac{(\nu - \nu_c)}{\nu_0(T)} \right] \,,
\ee
where
\be \label{fillingfractiondef}
 \nu := \left| \frac{\rho}{B} \right|
\ee
is the filling fraction and $\nu_c$ is the filling fraction at the critical field. The phenomenological function $\nu_0(T)$ is consistent with a power law down to very small temperatures, below which deviations from a power are seen \cite{Shahar}. In particular, if the ohmic resistivity is compared at equidistant points on opposite sides of the critical magnetic field, with distance measured by filling fraction, $\nu$, then eq.~\pref{phenodualform} implies
\be
 \rho_{xx}(\nu_c - \Delta \nu) = \frac{\rho_{\star\,xx}^2}{\rho_{xx}(\nu_c + \Delta \nu)} \,.
\ee

More remarkably, this duality also appears to hold beyond the linear-response regime. This is shown in Fig.~\ref{Fig:dual}, whose left panel plots the entire current-voltage relation for the corresponding points on either side of the critical point. Curves equidistant from the critical point (measured using filling fraction) are mirror images of one another, reflected through the line $V = I$. This is shown in the right panel, in which the upper curves are reflected and superimposed on the lower curves. This reflection invariance implies the relation $\rho_{xx} \to 1/\rho_{xx}$ when restricted to the slope of the approximately straight lines near zero voltage, which is the linear-response regime. But the figure shows it also applies in the regime for which $I(V)$ is noticeably curved. The full nonlinear reflection symmetry is equivalent to the condition $\rho_{xx}(V) \to 1/\rho_{xx}(V)$, where $\rho_{xx}(V) := \exd I/\exd V$ is the nonlinear, potential-dependent, resistivity.

\vspace{4mm}\noindent
{\em Super-universality:} Historically, the first evidence for duality came from the study of scaling behaviour as the temperature is lowered for magnetic fields chosen to lie at the transition between two plateaux (for a review, see {\em e.g.} \cite{ScalingRev}). The scaling occurs in the slope of the inter-plateau step in the Hall resistivity, which diverges in the zero-temperature limit. The width, $\Delta B$, of the region of nonzero ohmic resistivity between the two plateaux also scales, in that it vanishes like a power of temperature:
\be \label{rhoscaling}
 \frac{\exd \rho_{xy}}{\exd B} \propto T^{-\alpha}
 \quad \hbox{and} \quad
 \Delta B \propto T^\beta \,.
\ee
Remarkably, measurements not only show $\alpha = \beta = 0.42 \pm 0.01$ \cite{Wei} for the transition between two specific plateaux; they also show that the values of $\alpha$ and $\beta$ are the same for the transitions between different pairs of plateaux \cite{Wei}. This equivalence of scaling exponents for different transitions is called `super-universality', and is seen in Fig.~\ref{Fig:SuperU}. A nontrivial check on the AdS/CFT picture described below is its ability to account for this kind of scaling and these observed values for $\alpha$ and $\beta$.

\FIGURE[ht]{\includegraphics[scale=0.5]{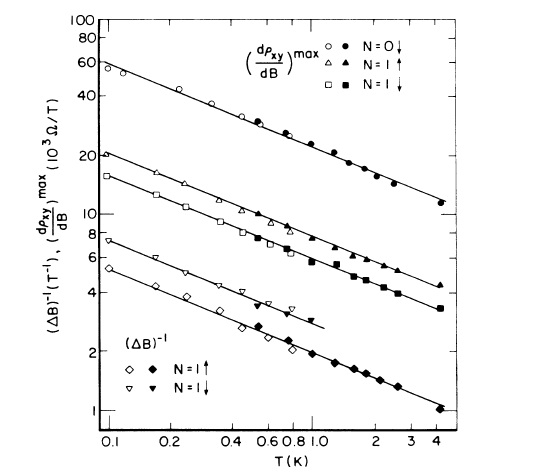}
\caption{Evidence for the super-universality -- the sharing of scaling exponents for transitions between different plateaux, reproduced from ref.~\cite{Wei}.
} \label{Fig:SuperU} }

\subsubsection*{Connection to duality}

What is not yet clear is why these striking observational features are evidence for duality.

Historically, early indications for duality in interacting systems \cite{Cardy} combined with the observed equivalence of scaling behaviour at the transitions between different critical points, together with the shape (in the conductivity plane) of the flow to low temperature to motivate the guess that a duality group might be relevant to quantum Hall systems. Early observations about duality \cite{Cardy} in field theory, and the similarity between the phase structure seen in the temperature flows and properties of $SL(2,Z)$ led the authors of ref.~\cite{LutkenRoss} to propose the existence of a group of symmetries acting on the complex conductivity $\sigma = \sigma_{xy} + i \sigma_{xx}$ (in units of $e^2/h$) according to
\be \label{gammadef}
    \sigma\rightarrow \frac{a\sigma + b}{c\sigma
    +d} \,,
\ee
where the integers $a$ through $d$ satisfy the constraint $ad-bc = 1$. It was subsequently noticed \cite{LutkenRossII,Lutken} that odd-denominator plateaux are singled out as endpoints to the temperature flow if the group is restricted to the subgroup $\Gamma_0(2)$ defined by the condition that the integer $c$ must be even,\footnote{In terms of the generators $S$ and $T$ of $SL(2,Z)$, defined below, $\Gamma_0(2)$ can be regarded as that subgroup generated by $ST^2S^{-1}$ and $T$.} leading to predictions for the universal values for the conductivities, like $\rho_{\star\,xx}$, at the critical points.

Similar conclusions were reached at much the same time in the condensed-matter community \cite{KLZ}, where more detailed thinking about the microscopic dynamics led to the Law of Corresponding States, whose action on filling fractions implies an action on conductivities of the $\Gamma_0(2)$ form. Once restricted to zero temperature these can be regarded as a set of transformations relating the ground state wave-functions for the various quantum Hall plateaux, as was implicit in the work of Jain and collaborators \cite{Jain}. Although the concrete connection of the experiments to what the electrons are doing was a step forward, a downside was the necessity to resort to mean-field reasoning (see however \cite{KF}).

\FIGURE[ht]{\includegraphics[scale=0.6]{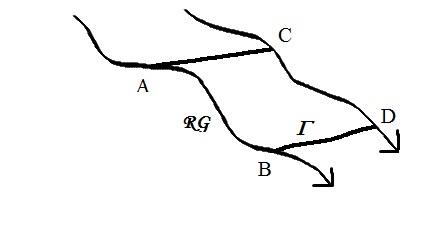}
\caption{The relation between RG flow and the action of the duality group, in the conductivity plane. If A flows to B, and D is B's image under the group $\Gamma$, then the RG flow must take C to D if C is A's image under $\Gamma$.
} \label{Fig:Commute} }

The precise relation between the above observations and a duality group came with the observation that {\em all} of the above experiments --- including the semi-circle law \cite{Semicircle}, universal critical points for transitions between general plateaux \cite{ModularSymmetry}\footnote{Spin effects can also modify the precise position of the critical points \cite{Gamma2,Huang}.}  and the validity of $\rho_{xx} \to 1/\rho_{xx}$ duality, even beyond linear response \cite{Nonlinear} --- follow as exact consequences of particle-hole invariance together with the assumption that the $\Gamma_0(2)$ action commutes with the RG flow of the conductivities in the low-energy theory. (Fig.~\ref{Fig:Commute} illustrates what it means for the action of the group to commute with the RG flow, and Fig.~\ref{Fig:Flowlines} shows a pattern of flow lines that is consistent with commuting with the duality group $\Gamma_0(2)$.)

Furthermore, there are good reasons to believe that such duality transformations, acting on the conductivities as in eq.~\pref{gammadef}, should actually arise in low-energy systems in two spatial dimensions. This was first argued \cite{PVD} as a general consequence of the similar kinematics of weakly interacting pseudo-particles and vortices, in a picture (like the `composite boson' framework, described below) where these were the dominant charge carriers in the low-energy effective theory.\footnote{Because this argument only relies on using duality to relate the conductivity produced by a vortex with that produced by a quasi-particle --- as opposed to trying to explicitly compute either result separately, as done in $\cite{KLZ}$ --- it can apply equally well at zero- and finite-temperature and so side-steps the objection of $\cite{HKSS}$ based on the subtleties of the ordering of the $T\to 0$ and $\omega \to 0$ limits.} In this language the two independent generators of $\Gamma_0(2)$ turn out to be 
particle-vortex duality \cite{PVDFL}, and the freedom to add $2\pi$ statistics flux to any quasi-particles.

Similar arguments showed that it would be a slightly different subgroup of $SL(2,Z)$ --- the subgroup\footnote{This subgroup is generated by the elements $S$ and $T^2$ of $SL(2,Z)$.} $\Gamma_\theta(2)$ --- that would be relevant to quantum Hall systems built from microscopic bosons rather than fermions \cite{PVD}. Because this group differs in detail from $\Gamma_0(2)$, it leads to the prediction of a suite of experimental results for bosonic quantum Hall systems that are similar to those described above (such as by including a semi-circle law), but which differ in detail (such as by predicting different plateaux)\footnote{$\Gamma_\theta(2)$ can also have implications for quantum Hall effects in more complicated systems, like graphene, where there is more than one species of conduction electron \cite{graphene}.} \cite{PVD}. In particular, the bosonic subgroup $\Gamma_\theta(2)$ contains the weak-strong duality transformation, $\sigma \to - 1/\sigma$, that is not present for the observed quantum Hall systems, 
but which was observed early on to be a symmetry of scalar electrodynamics in 2+1 dimensions \cite{ShapereWilczek}.

\FIGURE[ht]{\includegraphics[scale=0.5]{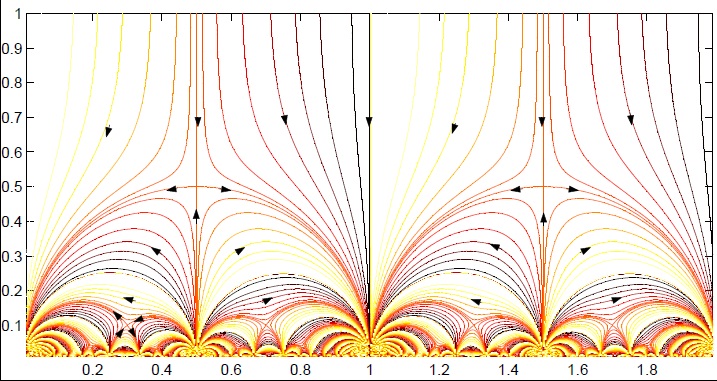}
\caption{A plot of some of the flow lines (for decreasing temperature) for the conductivities that are dictated by $\Gamma_0(2)$ invariance. The vertical axis represents $\sigma^{xx}$ and the horizontal axis is $\sigma^{xy}$ (in units of $e^2/h$). Flows are attracted to odd-denominator fractions at zero temperature, with bifurcations between different domains of attraction at specific magnetic fields. Notice that the semicircles that describe flow at constant magnetic field at the bifurcation between two basins of attraction are also lines along which the system moves when magnetic fields are varied at vanishingly small temperatures (colour online).
} \label{Fig:Flowlines} }

It has since been argued \cite{Witten} that eq.~\pref{gammadef} should emerge on very general grounds for {\em any} 2+1 dimensional CFT having a conserved $U(1)$ symmetry, making its emergence at low energies essentially automatic for any system having such a CFT governing its far-infrared behaviour. In particular, ref.~\cite{Witten} shows that it is the full $SL(2,Z)$ group that generically emerges in this way for theories defined in geometries that admit a spin structure, while only the subgroup $\Gamma_\theta(2)$ emerges if a spin structure is absent.

\subsection{The low-energy picture}

The overall picture that emerges from the convergence of theory and experiments for quantum Hall systems is as follows. In two spatial dimensions the huge degeneracy of Landau levels in a magnetic field leads to ground states that can be very sensitive to electron interactions, allowing the possibility of the strongly correlated Laughlin ground states describing the various quantum Hall plateaux. Transport properties near these plateaux at the low temperatures relevant to the conductivity measurements is governed by a low-energy effective theory obtained by integrating out the short-distance electron modes.

\subsubsection*{Far infrared: Integer quantum Hall systems}

In the very far infrared the effective zero-temperature theory obtained by integrating out all of the high-energy excitations is a function of the electromagnetic probe field, $A_\mu$, used to explore the electromagnetic transport:
\be \label{GammaFarIR}
 \Gamma_{\rm IR} = - \frac{k}{2\pi} \, e^2 \int_X \exd^3x \; \epsilon^{\mu\nu\lambda} A_\mu \partial_\nu A_\lambda \,,
\ee
where the electron charge, $e$, is temporarily restored, and $X$ denotes the region containing the quantum Hall fluid. Topological considerations \cite{Witten} imply the coefficient $k$ is in general quantized to be an integer.\footnote{Given a spin structure $k$ could be half-integer, however we take the case of no spin structure because for the quantum Hall experiments of most interest the Zeeman splitting is larger than the Landau level spacing. See however \cite{GirvinReview} for a review of more complicated cases where electron spins can be important, and \cite{Gamma2,bilayers} for preliminary discussions of how duality arguments change in this case.} The current arising from the probe field $A_\mu$ inferred from eq.~\pref{GammaFarIR} is
\be
 J^\mu = \frac{\delta \Gamma_{\rm IR}}{\delta A_\mu}
 = - \frac{k e^2}{2\pi} \epsilon^{\mu\nu\lambda} F_{\nu\lambda} \,,
\ee
which when evaluated with only $E_x = F_{tx}$ nonzero and compared with $J_i = \sigma_{ij} E_j$ implies the conductivities
\be
 \sigma_{xx} = 0 \quad \hbox{and} \quad
 \sigma_{xy} = - \sigma_{yx} = k \,,
\ee
in units of $e^2/h = e^2/2\pi$ (using $\hbar = 1$). Thus is captured the integer quantum Hall plateaux.

A potential puzzle about the low-energy action $\Gamma_{\rm IR}$ is that it is not gauge invariant when $X$ has a boundary, as real quantum Hall systems do. In this case the failure of gauge invariance in eq.~\pref{GammaFarIR} is canceled by a related failure coming from degrees of freedom that live exclusively on the boundary, $\partial X$. These degrees of freedom are the ones that actually transport the charge in the low-energy theory, which moves along the boundaries of the quantum Hall domains. Because these are restricted to the boundaries they are described by a chiral 1+1 dimensional CFT, whose $U(1)$ anomaly provides the required cancelation.

\subsubsection*{Far infrared: Fractional quantum Hall systems}

Another puzzle about eq.~\pref{GammaFarIR} is that the quantization of $k$ seems to preclude on general grounds the possibility of having fractional quantum Hall plateaux. A resolution to this puzzle is suggested by the `composite boson' picture of quantum Hall systems, as is now described \cite{Zhang}.\footnote{The related `composite fermion' model \cite{Heinonen} is widely used in theoretical studies of quantum Hall systems, and has also been discussed within an AdS/CFT framework \cite{SJR}.}

The composite boson model starts with the observation that statistics is a supple concept in 2+1 dimensions where fractional statistics are allowed, and in particular can be explicitly implemented through the artifice of having particles carry with them flux tubes of a fictitious electromagnetic field, $a_\mu$ \cite{anyonreview}. Specifically, if $S[\psi,A]$ is the action for point particles, $\psi$, having charge $e$ coupled to an electromagnetic field, $A_\mu$, then the deformation
\be
 S_\vartheta[\psi,A,a] := S[\psi,A+a] - \frac{e^2}{2\vartheta} \int \exd^3 x \; \epsilon^{\mu\nu\lambda} a_\mu \partial_\nu a_\lambda \,,
\ee
describes the same theory where the statistics of the $\psi$ particles is shifted by the angle $\vartheta$. For instance, if a two-particle state described by the action $S[\psi,A]$ originally acquired a phase $\eta$ when the two particles are interchanged, then when described by $S_\vartheta[\psi,A,a]$ they instead acquire the phase $\eta e^{i\vartheta}$ on interchange. They do so because the gaussian integral over $a_\mu$ produces a saddle point that sets its magnetic field, $b = \partial_x a_y - \partial_y a_x$, proportional to the charge density, which is nonzero where the particles are but vanishes where they are not. For point particles this is equivalent to attaching a flux quantum to each particle, and it is the Aharonov-Bohm phase of this flux that produces the change in statistics.

With this in mind, the electrodynamics of 2+1 dimensional fermions can instead be regarded as that of bosons coupled to a statistics field with angle
\be \label{fthetachoice}
 \vartheta = (2n+1)\pi \,.
\ee
In this picture the quantum Hall plateaux with fractions $1/(2n+1)$ can be qualitatively understood using the following mean-field picture. For a macroscopic number of bosons, the accumulated statistical flux can be thought of as a constant background field, $b$. But because the charge carriers couple only to the sum $A_\mu + a_\mu$, special things can happen when the real magnetic field cancels this background statistics field. For these special values where $B + b = 0$ the bosons see no net field, and so are free to Bose-Einstein condense --- producing a superconducting phase. This condensation is how the strongly correlated fractional quantum Hall state is understood in this picture. Due to the choice, eq.~\pref{fthetachoice}, the cancelation happens when the filling fraction is $\nu = 1/(2n+1)$, corresponding to the principle series of fractional states described by the Laughlin wave-function.

In this picture there is also a qualitative understanding of the stability of these plateaux to small changes of $B$. The `superconductor' then sees a net magnetic field, but the idea is that the superconductor is a Type II superconductor for which this field penetrates as a vortex without destroying the condensation. These vortices have fractional statistics, and correspond to the quasi-particles of the Laughlin fluid. The plateau ends for fields, $B$, large enough that there are so many vortices that the superconductivity is ruined. The picture then is that the vortices themselves condense, producing a quantum Hall state, $p/q$ with $p \ne 1$. This process continues generating the many plateaux observed in a hierarchical way \cite{Jain}. Although the mean-field arguments are suspect, this is a conceptually attractive framework for understanding quantum Hall dynamics, for which notions of particle-vortex duality are likely to be useful \cite{PVD}.

Coming back to the far-infrared effective action, the above picture suggests that eq.~\pref{GammaFarIR} should be generalized to
\bea \label{GammaFarIRTheta}
 \exp \Bigl\{ i\Gamma_{\rm IR}[A] \Bigr\} &:=&
 \int \cD a_\mu \;
 \exp \left\{- \frac{k e^2}{2\pi} \int_X \exd^3x \,
  \epsilon^{\mu\nu\lambda} (A_\mu + a_\mu) \partial_\nu
  (A_\lambda + a_\lambda) \right. \nn\\
 && \qquad\qquad\qquad\qquad \left.
 - \frac{e^2}{2\vartheta} \int \exd^3 x \; \epsilon^{\mu\nu\lambda} a_\mu \partial_\nu
 a_\lambda  \right\}  \,.
\eea
If the first term is the result that would be obtained, as above, from a system of electrons, then electrons could also give eq.~\pref{GammaFarIRTheta} for $\vartheta = 2n \pi$, since any shift of statistics by an integer multiple of $2\pi$ has no effect. Integrating out $a_\mu$, leads to the Hall conductivity
\be \label{fermionplateaux}
 \sigma_{xy} = \frac{k}{2nk + 1} \,,
\ee
which is a fraction (in units of $e^2/h = e^2/2\pi$), though always with an odd denominator.

For future reference, notice that a quantum Hall system built from bosons would instead correspond to the choice $\vartheta = (2n+1)\pi$, leading to
\be \label{bosonplateaux}
 \sigma_{xy} (\hbox{bosons}) = \frac{k}{(2n+1)k + 1}  \,.
\ee
In units of $e^2/h = e^2/2\pi$ this is a fraction $p/q$, with $q$ odd if $p$ is even, and vice versa. Note in particular that if all else is equal, then shifting statistics angle by $\vartheta \to \vartheta + \pi$ shifts the complex conductivity by\footnote{In terms of the generators $S(\sigma) = -1/\sigma$ and $T(\sigma) = \sigma + 1$, this corresponds to $\sigma \to ST^{-1}S(\sigma)$.}
\be \label{addflux}
 \frac{1}{\sigma} \to \frac{1}{\sigma} + 1 \,.
\ee

\subsubsection*{Not quite so deep in the infrared}

The interest in this paper is in the approach to the quantum Hall plateaux for small temperatures, rather than in the ground states themselves, and so the goal is to obtain an effective low-energy description that is not quite so far in the infrared as the Chern-Simons action just described. It is for this effective theory that any emergent duality group should be found if it is to be relevant for the experiments that probe the approach to, and transitions between, different quantum Hall plateaux.

The observational evidence is that this regime is described by some system with a $\Gamma_0(2)$ duality group that commutes with its RG flow, but real progress in constructing candidate effective field theories has been blocked by the lack of examples of strongly correlated systems explicitly displaying the emergent duality. Once such a model is in hand its implications that go beyond implications of duality can be tested, to see if it describes the experimental systems.

The remainder of this paper identifies a first candidate using recently developed tools from the AdS/CFT correspondence. As discussed in the introduction, for the present purposes, the great virtue of this correspondence is twofold: it provides a calculable laboratory of strongly interacting 2+1 dimensional systems; and it naturally produces systems having emergent duality groups.

\section{Holographic duality}

AdS/CFT formulations of 2+1 dimensional CFTs involve electromagnetic gauge fields in 3+1 dimensional asymptotically AdS backgrounds. Particle-vortex interchange in the CFT corresponds to the interchange of electric and magnetic fields on the AdS side, so part of the ease of having an emergent duality in the CFT is the propensity on the AdS side for the electromagnetic theory to be invariant under electric-magnetic interchange. Since this transformation takes the electromagnetic coupling from weak to strong (and vice versa), on the AdS side it is useful to have a scalar field, $\phi$, whose value tracks the size of this coupling. Here we use the modular symmetry as an input to constrain the model and do not derive it as an emergent symmetry.

Another generator is needed to obtain a group like $SL(2,Z)$ --- or one of the level-two subgroups, like $\Gamma_0(2)$ or $\Gamma_\theta(2)$ --- and given the above discussion it is natural to seek this as the freedom to change particle statistics by $2\pi$. Since particle statistics are described by a Chern-Simons term in the CFT, on the AdS side it is natural to seek a symmetry that shifts the coefficient of $F \wedge F$. For this reason it is also useful to have a scalar field, $\chi$, whose value tracks this interaction.

The minimal set of fields to follow in the AdS formulation should then be gravity, the electromagnetic field, plus the two scalars: the dilaton, $\phi$, and axion, $\chi$. These fields naturally appear in the low-energy limit of string theory, so the kinds of theories entertained here are likely to arise generically in more explicit string constructions. (In this paper we take a phenomenological point of view, and do not try to embed the 3+1 dimensional field theory into an explicit stringy framework. Although this would be instructive, most of the additional bells and whistles live at very high energies and so are likely to decouple from the low-energy limit that is always of interest for the applications we have in mind.)

The holographic interpretation of black holes with this field content has recently been worked out \cite{dilaton1,dilaton2}. Although these models cannot themselves directly provide descriptions of quantum Hall systems, since for nonzero magnetic fields their DC ohmic conductivity vanishes at finite temperature, they are interesting in their own right. This section briefly recaps some of their features, with the goal of describing the duality transformations of interest for the model of real interest in the next section.

\subsection{Maxwell and the axio-dilaton}

The starting point is the Einstein-Maxwell action coupled to the axio-dilaton in 3+1 dimensions:\footnote{We use a `mostly plus' metric signature and Weinberg's curvature conventions \cite{Wbg}, which differ from those of MTW \cite{MTW} only by an overall sign in the Riemann tensor.}
%
%
\bea \label{MaxwellDilatonAction}
 S &=& - \int \exd^4 x \, \sqrt{-g} \;
 \left\{ \frac{1}{2\kappa^2} \left[ R - 2\Lambda
 + \frac{\lambda^2}{2} \left( \partial_\mu \phi \,\partial^\mu \phi
 + e^{2\phi} \; \partial_\mu \chi \, \partial^\mu \chi
 \right) \right] \right. \nn\\
 && \qquad\qquad\qquad\qquad\qquad\qquad\qquad\qquad
  \left. + \frac14 \, e^{- \phi}
  F_{\mu\nu} F^{\mu\nu} + \frac14 \, \chi \, F_{\mu\nu} \Fd^{\mu\nu} \right\} \,,
\eea
where $\Fd_{\mu\nu} := \frac12 \, \epsilon_{\mu\nu\lambda\rho} F^{\lambda\rho}$, and $\epsilon_{\mu\nu\lambda\rho}$ has a factor of $\sqrt{-g}$ extracted so that it transforms as a tensor (rather than a tensor density). The constant $\Lambda = 3/L^2$ is the AdS cosmological constant and $\kappa^2 = 8 \pi G$ is Newton's constant, so weak curvature requires $\kappa^2/L^2 \ll 1$. Similarly, the Maxwell coupling is $g^2 \propto e^\phi$ so weak coupling corresponds to $e^\phi \ll 1$. The dimensionless parameter\footnote{We thank Elias Kiritsis for emphasizing the importance of this parameter, which for known supersymmetric examples satisfies $\lambda=1$.} $\lambda$ is at this point arbitrary, and can be absorbed by choosing $\hat\phi := \lambda \phi$ at the cost of re-appearing within the exponents: $e^{\phi} = e^{\hat\phi/\lambda}$.

\subsection{Duality relations}

The couplings of this action are chosen to ensure the existence of a duality group, and at the classical level there is an embarrassment of riches since the equations of motion are invariant under the group $SL(2,R)$. To see the action of this group define the axio-dilaton by
\be
 \tau := \chi + i e^{-\phi} \,,
\ee
for which weak coupling corresponds to large Im $\tau$. Then the $\chi$ and $\phi$ kinetic terms become
\be
 \partial_\mu \phi \, \partial^\mu \phi +
 e^{2\phi} \; \partial_\mu \chi \, \partial^\mu \chi
 = \frac{ \partial_\mu \tau \, \partial^\mu \overline \tau}{
 (\hbox{Im} \, \tau)^2} \,,
\ee
which is invariant under the transformations
\be \label{SL2Rtau}
 \tau \to \frac{a \, \tau + b}{c \, \tau + d}
 \quad \hbox{and} \quad
 g_{\mu\nu} \to g_{\mu\nu} \,,
\ee
where $a$, $b$, $c$ and $d$ are arbitrary real numbers that satisfy the $SL(2,R)$ condition $ad - bc = 1$.

To define the action on the Maxwell field, following \cite{GiRa} define
\be
 G^{\mu\nu} := - \frac{2}{\sqrt{-g}} \, \left( \frac{\delta
 S}{\delta F_{\mu\nu}} \right) = e^{-\phi} F^{\mu\nu} + \chi \Fd^{\mu\nu} \,,
\ee
which takes the simple form
\be \label{constitutivereln}
 \cG^{\mu\nu} = \overline \tau \, \cF^{\mu\nu} \,,
\ee
when written in terms of the complex quantities
\be
 \cF_{\mu\nu} := F_{\mu\nu} - i \Fd_{\mu\nu}
 \quad \hbox{and} \quad
 \cG_{\mu\nu} := - \Gd_{\mu\nu} - i G_{\mu\nu} \,.
\ee
Eq.~\pref{constitutivereln} is invariant under the transformation, eq.~\pref{SL2Rtau}, provided the Maxwell field transforms as
\be \label{Maxwelltransfn}
 \left( \begin{array}{c} \cG_{\mu\nu} \\
     \cF_{\mu\nu} \\ \end{array} \right)
 \to
 \left( \begin{array}{cc}
     a & b \\
     c & d \\
   \end{array} \right)
 \left( \begin{array}{c} \cG_{\mu\nu} \\
     \cF_{\mu\nu} \\ \end{array} \right) \,,
\ee
Since the Maxwell equations and the Bianchi identity are
\be
 \nabla_\mu \hbox{Im}\, \cG^{\mu\nu}
 = \nabla_\mu \hbox{Im}\, \cF^{\mu\nu} = 0 \,,
\ee
these are also invariant under $SL(2,R)$. The Maxwell contribution to the axio-dilaton equation is similarly invariant \cite{GiRa}.

\subsection{From $SL(2,R)$ to $SL(2,Z)$}

Although $SL(2,R)$ is a larger group than bargained for, in string theory it is generically only an artefact of the classical approximation, and is broken down to a discrete subgroup by quantum effects. Since the quantum plateaux ultimately prove to be in a strongly coupled part of parameter space (over which the unbroken discrete symmetries ultimately give calculational access -- see below), their properties are strongly affected by the breaking.

The low energy supergravity of Type IIB string theory has an action in 10 dimensions that is similar to the one described above, whose equations of motion are $SL(2,R)$ invariant. In this  case the symmetry is broken by the presence of objects whose charges are quantized. For example, a $(m,n)$-string ({\em i.e.} a bound state of a fundamental F-string with charge $m$ with a D-string with charge $n$)\footnote{We use $(m,n)$ rather than the more traditional $(p,q)$ to avoid notational conflict with our later use of $p$ and $q$.} has tension,
\begin{equation}
 \tau_{m,n} = e^\phi ( m + \chi n)^2 + e^{-\phi} n^2 \,.
\end{equation}
Under $SL(2,R)$ transformations, the $(m,n)$-string transforms into a $\left( m',n' \right)$-string, where
\be
 m^{'} = d m + c n \,, ~~~
 n^{'} = b m + a n \,.
\ee
Because $m$ and $n$ are quantized $SL(2,R)$ is broken to $SL(2,Z)$.

For holographic applications similar considerations are very likely to apply. In particular, probing the CFT at finite temperature and density require studying the AdS theory in the presence of a charged (dilatonic) black hole. This becomes a dyonic black hole --- with both electric and magnetic charges, $Q_e$ and $Q_m$ --- if the CFT is probed in an external magnetic field. Although these black holes are usually studied in the classical limit, in principle the AdS/CFT duality is exact and so quantum effects can also be studied. In particular, the Dirac quantization conditions for magnetic monopoles should apply, requiring the electric and magnetic charges to be quantized relative to one another. In microscopic brane constructions, dyonic objects in the bulk can be identified as charged solitons in the boundary CFT \cite{HY}.

It then suffices that there should be a minimum electric charge to learn that magnetic and electric charges must be quantized in terms of this minimum charge. As we see below, such a quantization on the AdS side naturally leads to a quantization of the Hall conductivities on the CFT side: $\sigma_{xy} \sim Q_e/Q_m \sim p/q$, for integer $p$ and $q$. The precise pattern of fractions that is allowed depends on the precise discrete subgroup --- possibly $SL(2,Z)$, $\Gamma_0(2)$ or $\Gamma_\theta(2)$ --- of $SL(2,R)$ that is left unbroken by the full string dynamics. Since several specific stringy ultraviolet completions are likely to exist for the given low-energy action, eq.~\pref{MaxwellDilatonAction}, and since different systems give rise to different discrete symmetries \cite{SW}, in the phenomenological approach followed here we imagine ourselves to be free to choose this unbroken discrete symmetry.

\subsection{Conductivities}

Computing the ohmic and Hall conductivities as functions of temperature, charge density and magnetic field requires studying the response of the above AdS system to small electromagnetic perturbations about a dyonic axio-dilaton black hole. This is explored in some detail in refs.~\cite{dilaton1,dilaton2}.

\subsubsection*{Action of $SL(2,R)$}

In particular, these authors compute the action of the underlying $SL(2,R)$ symmetry on the conductivities, and show that they take the form of eq.~\pref{gammadef}. We reproduce a version of the argument here that generalizes easily to the case of later interest.

The starting point is the AdS/CFT translation table,\footnote{There is generally a choice of CFT, depending on the precise form of the boundary conditions used in AdS \cite{BF,KW}. In the present instance ref.~\cite{Witten} argues that one of these choices can be regarded as equivalent to treating the gauge field on the boundary as dynamical, as would be done when coupling to a statistics field in 2+1 dimensions. Furthermore, such choices are implicitly made when comparing theories related by transformations involving $S$-duality, $\tau \to -1/\tau$. These complications do not play a direct role in what follows.} which gives the electromagnetic current, $J^a$, when the CFT is perturbed by an electromagnetic field, $F_{ab}$. On the AdS side the perturbation is obtained by solving the linearized Maxwell equation, and evaluating the action as a function of the perturbation on the boundary. Differentiating with respect to $A_\mu$ to get the current gives a simple form when expressed in terms of $G^{\mu\nu}$:
\be
 J^a = \left. \sqrt{-g} \; G^{v a} \right|_0 \,,
\ee
where $v$ is a radial coordinate ({\em i.e.} a function of $r$) for which conformal infinity lies at $v = 0$ and the horizon is at $v = v_h$.

Focusing on the spatial components, $J^x$ and $J^y$, and using the (real part of the) transformation rule eq.~\pref{Maxwelltransfn}, then implies
\be \label{Currenttransfn}
 \left( \begin{array}{c} \cJ \\
     \cE \\ \end{array} \right)
 \to
 \left( \begin{array}{cc}
     a & b \\
     c & d \\
   \end{array} \right)
 \left( \begin{array}{c} \cJ \\
     \cE \\ \end{array} \right) \,,
\ee
where\footnote{Our convention is $\epsilon^{tvxy} = +1/\sqrt{-g}$,
so is opposite to \cite{dilaton2}.}
\be
 \cJ := \left[ - \Gd_{tx} + i \Gd_{ty} \right]_0
 = \Bigl[ - \sqrt{-g} \Bigl( G^{vy} + i G^{vx} \Bigr) \Bigr]_0
 = -i \left( J^x -i J^y \right) \,,
\ee
and
\bea
 \cE := \left[ F_{tx} - i F_{ty} \right]_0 = E_x -i E_y\,.
\eea

But in linear response the conductivity tensor is defined\footnote{From this point on we adopt consistent tensor conventions for the conductivity, which is naturally contravariant.} to be $J^i = \sigma^{ij} \, E_j$, or equivalently (keeping in mind $\sigma^{yx} = - \sigma^{xy}$ and $\sigma^{xx} = \sigma^{yy}$ for rotationally invariant systems),
\bea
 \cJ = -J^y - i J^x &=&
 -\left(\sigma^{yx} E_x + \sigma^{yy} E_y \right) - i
 \left( \sigma^{xx} E_x + \sigma^{xy} E_y \right) \nn\\
 &=& -\left( \sigma^{yx} + i \sigma^{xx} \right)
 \left( E_x - i E_y \right)
 = \sigma_- \cE \,,
\eea
where $\sigma_- := \sigma^{xy} - i \sigma^{xx}$. Consistency of this relation with the transformation, eq.~\pref{Currenttransfn}, then implies
\be
 \sigma_- \to \frac{ a \sigma_- + b}{c \sigma_- + d} \,.
\ee
Complex conjugation -- we consider here only DC conductivities, whose imaginary parts vanish --- then also implies the desired transformation, eq.~\pref{gammadef}, for $\sigma = \sigma_+ = \sigma^{xy} + i \sigma^{xx}$.

\subsubsection*{Classical conductivities}

The authors of refs.~\cite{dilaton1,dilaton2} also show that the low-temperature properties of the conductivities predicted by this theory are relatively simple. The strategy is first to compute explicitly in the case of a purely electric black brane with a vanishing axion field. The general result for dyonic branes with an axion is then found by performing an appropriate $SL(2,R)$ transformation.

The appropriate black brane geometries have the form
\be
 \exd s^2 = - \mathfrak{h}^2(r) \exd t^2 + \frac{\exd r^2}{\mathfrak{h}^2(r)}
 + \mathfrak{b}^2(r) \, \Bigl( \exd x^2 + \exd y^2 \Bigr) \,,
\ee
for which the Maxwell field equation $\nabla_\mu G^{\mu\nu} = 0$ has solution
\be
 G^{rt} = - \,\frac{Q_e}{\mathfrak{b}^2(r)} \,,
\ee
and so using the constitutive relation, $G^{\mu\nu} = e^{-\phi} F^{\mu\nu} + \chi \Fd^{\mu\nu}$, then gives (with $F_{xy} = Q_m$)
\be
 F = (Q_e - \chi Q_m) \frac{e^\phi}{\mathfrak{b}^2} \;
 \exd r \wedge \exd t + Q_m \, \exd x \wedge \exd y \,.
\ee

Given the $SL(2,R)$ transformation rules for the Maxwell field, these expressions imply an action of $SL(2,R)$ on the charges $Q_e$ and $Q_m$. Our strategy is to start with an electric dilaton brane with unit electric charge, zero magnetic charge, $\phi = \hat\phi_0$ and $\chi=0$. $\hat \phi_0$ is then chosen so that this configuration is mapped into a more general configuration with $Q_e$, $Q_m$, $\phi=\phi_0$ and $\chi=\chi_0$.

The behaviour of the purely electric brane with no axion is simple because at low temperatures and frequencies it is governed by the near-horizon limit of the near-extremal geometry, which is \cite{Taylor}
\be \label{nearhorizon}
 \exd s^2 \approx - \frac{r^2}{l^2} \left[ 1 - \left(
 \frac{r_h}{r} \right)^{2\zeta +1} \right] \exd t^2
 + \frac{ l^2 \;\exd r^2}{r^2 [1 - (r_h/r)^{2\zeta + 1}]}
 + r^{2\zeta} \left( \exd x^2 + \exd y^2 \right) \,.
\ee
This benefits from an attractor mechanism \cite{attractormech, attractormechrev} that makes the near-horizon geometry independent of the boundary data for the scalar fields at infinity. This implies that the constants $l$ and $\zeta$ are determined by the field equations, leaving the position of the horizon, $r_h$, as the only important scale. The same geometry also describes the near-horizon limit when the dilaton-Maxwell action is replaced by the dilaton-DBI action discussed below (as is shown in Appendix D).

In particular, the prediction \cite{dilaton1,dilaton2} $\zeta = 1/(1 + 4 \lambda^2)$ --- which comes from solving the field equations for the $SL(2,R)$-invariant action given above, eq.~\pref{MaxwellDilatonAction} --- is likely to be significant because the geometry of eq.~\pref{nearhorizon} is Lifshitz-like, with different scaling assigned to time and space directions. This is true even though the asymptotic geometry near infinity is relativistic, due to the presence of the dilaton. The dynamical exponent predicted at low temperatures (in the IR) in this case is
%
%
\be
 z= \frac{1}{\zeta} = 1 + 4 \lambda^2 \,,
\ee
although the asymptotic value, $z = 1$, would continue to apply in the UV. To the extent that this metric also describes the near-horizon limit of the background geometry in DBI-based model discussed below, we choose $\lambda$ to ensure that $z$ is consistent with low-temperature observations of scaling exponents. Since these indicate\footnote{We thank E. Fradkin and S. Kivelson for pointing out the evidence for $z = 1$.} $z = 1$ \cite{QHEz=1sc}, as is also suggested by the importance of Coulomb physics in the microscopic picture \cite{QHEz=1th}, in practice we imagine taking $\lambda^2 \ll 1$, although we expect that the classical approximation to break down for sufficiently small $\lambda$. By contrast, the supersymmetric choice $\lambda = 1$ predicts $z = 5$.

The dilaton also varies logarithmically with $r$ in the purely electric solution, $e^{\phi} \propto r^{4\zeta}$, which vanishes on the horizon in the extremal case ($r_h \to 0$). For magnetic branes ($Q_e = 0$ and $Q_m \ne 0$) the dilaton is instead driven to the strong-coupling regime at the horizon in the extremal case. Control is nonetheless maintained in refs.~\cite{dilaton1,dilaton2} by taking $T$ to be nonzero but small, so the brane is not quite extremal. Then an asymptotic value for the dilaton at conformal infinity can be chosen to ensure that the coupling remains weak enough right down to $r = r_h \ne 0$. This tendency to strong coupling at low enough temperatures (for fixed dilaton) is an important feature of these dual systems, that in later sections also limits our ability to compute conductivities directly near quantum Hall plateaux using semiclassical methods. (It is recourse to the unbroken discrete symmetries, like $SL(2,Z)$, that ultimately allow progress nonetheless.)

The explicit form obtained in this way for the AC conductivities in the limit $\omega \ll T \ll \mu$ (where $\mu$ is the chemical potential required to maintain a charge density $\rho \sim Q_e$) is \cite{dilaton2}
\be
 \sigma_{xy} = \frac{\rho}{B} \Bigl[1 + \cO \left( \omega^2
 \right) \Bigr] \,,
 \quad \hbox{and} \quad
 \sigma_{xx} = \cO \left( \omega \right) \,.
\ee
In particular, there is no DC ohmic conductivity. This ultimately vanishes because the ohmic conductivity is infinite at zero $B$ due to translation invariance \cite{dilaton2}. Although $SL(2,R)$ is nicely realized by the RG flow, $\exd \tau/\exd r$, of the axio-dilaton \cite{dilaton2}, it cannot directly describe the temperature flow of DC conductivities in quantum Hall systems.\footnote{Ref.~\cite{dilaton2} also models DC conductivity due to disorder by giving the frequency a small imaginary part.} For this reason we next explore a slightly more complicated system for which $SL(2,R)$ invariance coexists with nonzero DC conductance.

\section{Quantum Hall-ography}

In order to obtain DC conductivity in an $SL(2,R)$ invariant way, we follow ref.~\cite{StrangeM} and study the case of a probe brane, described by the DBI action, situated within the background geometry of an appropriately chosen black brane. As discussed in \cite{StrangeM}, the probe limit is crucial for obtaining DC ohmic resistance because the infinite bath represented by the black brane can provide the required dissipation. Ideally, one would prefer not to have to rely on the probe approximation to achieve DC resistance, such as by incorporating disorder or some other breaking of translation invariance. We regard our reliance on the probe approximation here to be a temporary crutch that will not survive more sophisticated modeling.

\subsection{The setup}

The action for the revised model has the following form
\be
 S = S_{\rm grav} + S_{\rm gauge} \,,
\ee
where the gravitational sector is the same as before,
%
%
\be \label{EinsteinDilatonAction}
 S_{\rm grav} = - \int \exd^4 x \, \sqrt{-g} \;
 \left\{ \frac{1}{2\kappa^2} \left[ R - 2\Lambda
 + \frac{\lambda^2}{2} \left( \partial_\mu \phi \,\partial^\mu \phi
 + e^{2\phi} \; \partial_\mu \chi \, \partial^\mu \chi
 \right) \right] \right\} + S_{\rm Lifshitz} \,,
\ee
with the possible addition of a `Lifshitz' sector, whose purpose is to build in various features of the background geometry. For instance, in \cite{StrangeM} this sector is imagined to involve various Kalb-Ramond fields, $H_{\mu\nu\lambda}$, whose presence is used to generate an uncharged black-brane geometry that asymptotically scales spatial and temporal directions differently. The resulting asymmetric exponent $z=2$ was then chosen to achieve some strange-metal properties, like a resistivity linear in temperature.

Although not strictly necessary for quantum Hall plateaux, a similar construction could be used here to build in an arbitrary value of $z$ in the UV. What proves to be a more attractive choice, however, is instead to choose the Lifshitz sector such that its background metric is that of a dyonic black brane, whose extremal near-horizon geometry is that discussed in \S3, above, or its DBI generalization discussed in Appendix D. This is attractive because this automatically gives $z \simeq 1$ in the UV, while allowing $z$ to be dialed in the IR through the choice of the parameter $\lambda$. Potential sources for such a background are discussed below, after describing the gauge action, $S_{\rm gauge}$.

For the present purposes the main change relative to \S3\ is the gauge action, which replaces the dilaton-Maxwell form of eq.~\pref{MaxwellDilatonAction} with the DBI form
\begin{eqnarray}  \label{DBI-dilaton-axion}
 S_{\rm gauge} &=& - \cT \int \exd^4x \;
 \left[  \sqrt{-\det\left( g_{\mu\nu}
 +  \ell^2 \, e^{-\phi/2} F_{\mu\nu} \right)} - \sqrt{-g} \right]
 -\frac14 \, \int \exd^4x \, \sqrt{-g}
 \; \chi F_{\mu\nu} \Fd^{\mu\nu} \nn\\
 &=& - \cT \int \exd^4x \, \sqrt{-g} \; \left[ \sqrt{1 +
 \frac{\ell^4}{2} \, e^{-\phi} F_{\mu\nu} F^{\mu\nu} -
 \frac{\ell^8}{16} \,  e^{-2\phi} \left( F_{\mu\nu} \Fd^{\mu\nu}
 \right)^2}  - 1 \right] \nn\\
 && \qquad\qquad\qquad\qquad\qquad\qquad
  -\frac14 \, \int \exd^4x \, \sqrt{-g} \; \chi F_{\mu\nu} \Fd^{\mu\nu} \,, \end{eqnarray}
where the second line holds in 3+1 dimensions.

Eq.~\pref{DBI-dilaton-axion} is the unique $SL(2,R)$-invariant generalization of the DBI action \cite{GiRa}, and has the same form as would the action of a D3-brane written in Einstein frame if the quantity $\ell$ were given by
\be
 \ell^2 = 2 \pi \alpha'  \,,
\ee
with $\cT$ representing the brane tension. However, our approach here is phenomenological and nothing would change if this action were instead to emerge as the low-energy limit of some more complicated configuration involving other kinds of branes. Although we do not try to do so here, any full string embedding would require a precise statement of the position of the relevant branes in the extra dimensions, and of what stabilizes their motion (and gives mass to any other potentially light degrees of freedom). Presumably, the DBI action describes the dynamics of 2+1 D matter fields coupled with a strongly interacting CFT modeled by the background geometry. The matter fields are also coupled with the 3+1 D $U(1)$ gauge field on the probe brane. We imagine there to be a suitable large-$N$ limit in play, allowing us to neglect quantum fluctuations of fields on the AdS side.

This kind of dilaton-DBI action could also be used for the Lifshitz sector in the case where the background geometry is taken to be the near-horizon, near-extremal form described in \S3\ and Appendix D. If so, it would require a different $U(1)$ gauge potential and a parametrically larger tension $\cT \to \sim N \cT$ to justify the use of the probe approximation for the brane that produces the conductivity. It seems (and probably is) redundant to have the additional Lifshitz sector to produce such a background, when the same geometry would also be produced if $S_{\rm gauge}$ were treated beyond the probe approximation. We only do so here since we require the probe approximation in order to obtain a nonzero DC ohmic resistivity, and regard this as a feature to be improved in future iterations.

\subsection{Duality relations}

The important property of the DBI action used above is that it shares the duality invariance \cite{GiRa} of the dilaton-Maxwell action described earlier. The main change relative to the earlier discussion is the form of the constitutive relation between $G^{\mu\nu}$ and $F_{\mu \nu}$, which in this case is
\begin{eqnarray} \label{Gcalc}
 G^{\mu\nu} &=& - \frac{2}{ \sqrt{-g}} \,
 \left( \frac{\delta S}{\delta F_{\mu\nu}}\right) \nn\\
 &=& \frac{\cT \, \ell^4}{X} \;
 \left[ e^{-\phi} F^{\mu\nu}
 - \frac{\ell^4}{4} \, e^{-2\phi} \left( F_{\mu\nu}
 \Fd^{\mu\nu} \right) \Fd^{\mu\nu} \right]
 + \chi \Fd^{\mu\nu} \,,
\end{eqnarray}
where
\be
 X := \sqrt{1 +
 \frac{\ell^4}{2} \, e^{-\phi} F_{\mu\nu} F^{\mu\nu} -
 \frac{\ell^8}{16} \,  e^{-2\phi} \left( F_{\mu\nu} \Fd^{\mu\nu}
 \right)^2} \,.
\ee

In terms of this quantity gauge field equations and Bianchi identities have the same form as before,
\be
 \nabla_\mu G^{\mu\nu} = \nabla_\mu \Fd^{\mu\nu}
 = 0 \,.
\ee
It can be shown \cite{GiRa} that these --- and the other field equations and the constitutive relation, eq.~\pref{Gcalc} --- are invariant under the same $SL(2,R)$ transformations of the dilaton-Maxwell theory, eqs.~\pref{SL2Rtau} and \pref{Maxwelltransfn}:
\be
 \tau \to \frac{a \, \tau + b}{c \, \tau + d}
 \quad \hbox{and} \quad
 \left( \begin{array}{c} \cG_{\mu\nu} \\
     \cF_{\mu\nu} \\ \end{array} \right)
 \to
 \left( \begin{array}{cc}
     a & b \\
     c & d \\
   \end{array} \right)
 \left( \begin{array}{c} \cG_{\mu\nu} \\
     \cF_{\mu\nu} \\ \end{array} \right) \,,
\ee
with $g_{\mu\nu}$ fixed. As before $\cF_{\mu\nu} = F_{\mu\nu} - i \Fd_{\mu\nu}$ and $\cG_{\mu\nu} = - \Gd_{\mu\nu} - i G_{\mu\nu}$.

Because the symmetry acts in the same way on $G^{\mu\nu}$ as in the last section, the same conclusion is also true for the transformation laws for the current,
\be
 J^a = \Bigl. \sqrt{-g} \; G^{va} \Bigr|_0 \,.
\ee
It immediately follows that the conductivities of the dual CFT also transform as before, eq.~\pref{gammadef}:
\be
 \sigma \to \frac{a \, \sigma + b}{c \, \sigma + d} \,.
\ee

\subsubsection*{Beyond linear response}

The fact that the quantities $\cG_{\mu\nu}$ and $\cF_{\mu\nu}$ transform under $SL(2,R)$ in the same way as they did for the dilaton-Maxwell theory carries some potentially interesting implications. In particular, since the constitutive relation, eq.~\pref{Gcalc}, states that $G^{\mu\nu}$ is a linear combination of $F^{\mu\nu}$ and $\Fd^{\mu\nu}$ (with field-dependent scalar coefficients), it can always be written in a form similar to eq.~\pref{constitutivereln}:
\be
 \cG_{\mu\nu} = \overline \tau_{\rm eff} \, \cF_{\mu\nu} \,,
\ee
for some field-dependent quantity $\tau_{\rm eff} = \tau_{\rm eff} \left(\tau, F^2, F \cdot \Fd \right)$, satisfying $\tau_{\rm eff}(\tau,0,0) = \tau$. But the invariance of this relation under $SL(2,R)$ implies that the quantity $\tau_{\rm eff}$ must also transform under $SL(2,R)$ as
\be
 \tau_{\rm eff} \to \frac{ a \, \tau_{\rm eff} + b}{ c \,
 \tau_{\rm eff} + d} \,.
\ee
The quantity $\tau_{\rm eff}$ plays the role of a `dressed' axio-dilaton for the DBI theory.

A similar observation also holds for the quantities $\cJ = -i(J^x -i J^y)$ and $\cE = E_x - i E_y$ of the CFT. These inherit from $\cG_{\mu\nu}$ and $\cF_{\mu\nu}$ the same transformation as for the dilaton-Maxwell theory, \pref{Currenttransfn}:
\be
 \left( \begin{array}{c} \cJ \\
     \cE \\ \end{array} \right)
 \to
 \left( \begin{array}{cc}
     a & b \\
     c & d \\
   \end{array} \right)
 \left( \begin{array}{c} \cJ \\
     \cE \\ \end{array} \right) \,.
\ee
Defining the effective, field-dependent, conductivites, $\sigma^{xy}_{\rm eff}$ and $\sigma^{xx}_{\rm eff}$, by
\be
 \sigma_{{\rm eff}\,-} = \sigma^{xy}_{\rm eff} - i \sigma^{xx}_{\rm eff} := \frac{\cJ}{\cE} \,,
\ee
then implies that these must transform under $SL(2,R)$ as
\be
 \sigma_{{\rm eff}\,-} \to \frac{a \, \sigma_{{\rm eff}\,-}
 + b}{c \, \sigma_{{\rm eff}\,-} + d} \,,
\ee
and similarly for $\sigma_{\rm eff} := \sigma^{xy}_{\rm eff} + i \sigma^{xx}_{\rm eff}$.

We see here within an AdS/CFT realization how the implications of duality can apply beyond the strict linear-response regime, to include the nonlinear dependence of the conductivities on the applied fields. This is precisely what is required to account for some of the observations discussed in \S2\ (see Fig.~\pref{Fig:dual} and refs.~\cite{Nonlin, Nonlinear}).

\subsection{Holographic DC conductivities}

We next turn to the calculation of the conductivities as functions of temperature and magnetic field, to verify the presence of a nonzero DC ohmic conductivity.

\subsubsection*{Background geometry}

Following \cite{StrangeM} we take the background metric to solve the field equations generated only by $S_{\rm grav}$, and regard the effects of $S_{\rm gauge}$ as a perturbation to this geometry (the probe-brane approximation). We return below to the limitations of the domain of validity of this approximation.

We assume the background 4D geometry sufficiently near the black hole is
\be \label{metric}
 \exd s^2 = L^2 \left[ -h(v) \, \frac{\exd t^2}{v^{2z}}
 + \frac{\exd v^2}{v^2 h(v)} + \frac{\exd x^2
 + \exd y^2}{v^2} \right] \,,
\ee
where $L$ is the length scale defined by $\Lambda = 3/L^2$ (set to unity in what follows), and the Lifshitz parameter, $z$, measures the difference between the scaling dimension of the space and time directions, with $z = 1$ corresponding to equal scaling.\footnote{As discussed in \cite{StrangeM} the presence of $z$ complicates the discussion of the boundary conditions (see also footnote 11), particularly once $z \gsim 2$. Following \cite{Witten}, we expect these to be automatically incorporated into the duality transformations, but do not expect them to affect our conductivity calculations in any case.}

Not much is required to be known about the function $h(v)$, apart from that it is positive, approaches unity as $v \to 0$, and is assumed to have a simple zero, $h(v_h) = 0$ for $v_h > 0$, corresponding to the horizon of the black brane. The position of this horizon provides a temperature for the boundary theory in the usual way,
\begin{equation}
 T = \frac{|h'(v_h)|}{4\pi v_h^{z-1}} \sim \frac1{v_h^z} \,,
\end{equation}
with the approximate equality following from the assumption that $h'(v_h) \sim 1/v_h$. As before, the position of conformal infinity is taken to be $v = 0$.

If the black brane of the background geometry does not couple to a Maxwell field, as for the Lifshitz sector of ref.~\cite{StrangeM}, then the dilaton and axion fields can be taken to be constants: $\phi = \phi_0$ and $\chi = \chi_0$. In this case the parameter $z$ can be taken to be a knob to be dialed essentially at will. Alternatively, if the background geometry carries a charge and so approaches an extremal black brane at low temperature with an attractor form, then $\phi$ generically has a nontrivial profile. When necessary we take this to be 
\be
e^{\phi} \propto v^{-4} \label{dilatonraddep}
\ee
as suggested by the dilaton-Maxwell solution of \cite{Taylor,dilaton1} or the dilaton-DBI solution described in Appendix D. In either case the axion can be set to zero and then later regenerated by performing an $SL(2,R)$ transformation.

\subsubsection*{Conductivity calculation}

We proceed following closely the steps of ref.~\cite{StrangeM} (see also refs.~\cite{obannon,Drag}. The field equations for the gauge field are $\nabla_\mu G^{\mu\nu} = 0$, with $G^{\mu\nu}$ given by (\ref{Gcalc}). Those for the axio-dilaton are
%
%
\bea  \label{dilatoneom}
 0 &=& \lambda^2 \left[ \Box\phi -e^{2\phi}\partial_\mu\chi\partial^\mu\chi \right] + \frac{\kappa^2 \cT \ell^4}{2X}
 \left[ e^{-\phi} F_{\mu\nu} F^{\mu\nu}
 - \frac{\ell^4}{4} \, e^{-2\phi} \left( F_{\mu\nu} \Fd^{\mu\nu}
 \right)^2 \right]
  \nn\\
 &=& \lambda^2 \left[ \Box\phi  -e^{2\phi}\partial_\mu\chi\partial^\mu\chi \right] + \frac{\kappa^2}{2} \left[ G^{\mu\nu} - \chi \Fd^{\mu\nu}
 \right] F_{\mu\nu} 
\eea
and
\be \label{axioneom}
 \lambda^2 \nabla_\mu \Bigl( e^{2\phi} \, \nabla^\mu \chi \Bigr)
 - \frac{\kappa^2}{2} F_{\mu\nu} \Fd^{\mu\nu} = 0 \,.
\ee
The strategy in the probe limit is to solve the Maxwell equation, but neglect the corrections to the background metric and dilaton. The above equations show this requires the neglect of quantities like $\kappa^2 \cT/X$ and $\kappa^2 F_{\mu\nu} \Fd^{\mu\nu}$ relative to $1/L^2$ (which itself must satisfy $1/L^2 \ll 1/\ell^2$). Because $\kappa^2 \sim \ell^8/\Omega \ll \ell^2$ --- with $\Omega$ the volume of the extra dimensions not made explicit here --- these conditions need not imply that quantities like $\ell^4 e^{-\phi} F_{\mu\nu} F^{\mu\nu}$ are also small, so it remains consistent to keep the nonlinearities in the DBI action. In addition to these conditions are the more `stringy' conditions for weak coupling, $e^\phi \ll 1$, and the absence of runaway string pair-production \cite{Einstab} (more about the domain of validity later).

It suffices to compute the ohmic conductivity in the absence of a magnetic field and axion, since the general case can then be recovered by performing an appropriate $SL(2,R)$ transformation. To this end we require the solution to the Maxwell equation subject to the ansatz
\begin{equation}
  A = \Phi(v) \, \exd t + \Bigl[ \cA(v) - E t \Bigr] \exd x \,. \label{ohmicansatz}
\end{equation}
The corresponding components to the field strength then are
\bea
 &&F_{vt} = \Phi' \,, \quad
 F_{vx} = \cA' \,, \quad
 F_{tx} = E \nn\\
 \hbox{and} \quad
 &&\Fd^{xy} = - \frac{\Phi'}{\sqrt{-g}} \,, \quad
 \Fd^{ty} = \frac{\cA'}{\sqrt{-g}} \,, \quad
 \Fd^{vy} = - \frac{E}{\sqrt{-g}} \,,
\eea
and so $F_{\mu\nu} \Fd^{\mu\nu} = 0$.

Since the equations of motion can be written $\partial_\nu \left[ \sqrt{-g} \, G^{\nu\mu} \right] = 0$, the equations corresponding to $\mu = a = \{ x,y,t \}$ immediately integrate to give $\sqrt{-g} \, G^{va} = C^a$, where $C^a$ are three $v$-independent integration constants. The absence of an axion allows the choice $C^y = 0$, but the other two equations determine $\Phi'$ and $\cA'$ in terms of $C^t$ and $C^x$, as follows:
\be
 \sqrt{-g} \, \left( \frac{\cT \ell^4 e^{-\phi}}{ X} \right) \,
 g^{vv} g^{tt} \, \Phi' = C^t \quad
 \hbox{and} \quad
 \sqrt{-g} \, \left( \frac{\cT \ell^4 e^{-\phi}}{X} \right) \,
 g^{vv} g^{xx} \cA' = C^x \,,
\ee
where
\be
 X = \sqrt{ 1 + \ell^4 \,  e^{-\phi} \Bigl[ g^{vv} g^{tt} (\Phi')^2
 + g^{vv} g^{xx} (\cA')^2 + g^{tt} g^{xx} E^2 \Bigr] } \,.
\ee
Using these expressions to eliminate $\Phi'$ and $\cA'$ gives the following result for $X$ as a function of $C^t$ and $C^x$:
\be
 X = \sqrt{\frac{N}{D}} \,,
\ee
with
\bea
 N &:=& 1 + \ell^4 \, e^{-\phi} \left( \frac{E^2}{g_{tt} g_{xx}}
 \right) \nn\\
 D &:=& 1 + \frac{e^\phi}{\cT^2 \ell^4} \left[ \frac{(C^t)^2}{ g_{xx}^2}
 + \frac{(C^x)^2}{ g_{tt} g_{xx}} \right] \,.
\eea

Notice that when $v \to 0$ all of the metric functions diverge, and so both $N$ and $D$ approach unity. But when $v \to v_h$ we instead have $g_{tt} \to 0^-$ and $g_{vv} \to \infty$, while $g_{xx}$ and $\sqrt{-g}$ remain finite. This implies both $N$ and $D$ approach $- \infty$ in this limit, requiring they both change sign somewhere in the interval $0 < v < v_h$. A quick way to solve for the relation between $C^a$ and $E$ is the observation \cite{obannon} that the reality of the action requires both $N$ and $D$ to change sign at the same point, $v = v_\star$, implying
\be \label{vstareq1}
 -(g_{tt} g_{xx} )_\star = \frac{h(v_\star)}{v_\star^{2(z+1)}} =  \ell^4 \, e^{-\phi_\star} E^2 \,,
\ee
and
\be \label{vstareq2}
 -\frac{(C^x)^2}{(g_{tt} g_{xx})_\star} = \frac{(C^t)^2}{(g_{xx}^2)_\star}
 +  \cT^2 \ell^4 e^{-\phi_\star} \,.
\ee
The first of these can be used to infer the value of $v_\star$ as a function of $E$, and the second then imposes an $E$-dependent relation between $C^x$ and $C^t$. Notice that as $E \to 0$, eq.~\pref{vstareq1} implies $v_\star \to v_h \propto T^{-1/z}$.

Now the usual AdS/CFT translation tells us that the integration constants found above are the currents\footnote{A note on units of charge: this can be changed for the carriers in the CFT by rescaling $A_{\mu} \to \xi A_{\mu}$. This is a symmetry of the action --- contained in $SL(2,R)$ --- if $e^{-\phi} \to \xi^{-2} e^{-\phi}$ and $\chi \to \xi^{-2} \chi$. Under this rescaling $G^{\mu\nu} \to \chi^{-1} G^{\mu\nu}$, $J^\mu \to \xi^{-1} J^\mu$ and $\sigma^{ab} \to \xi^{-2} \sigma^{ab}$.} in the CFT: $J^a = C^a$, so using $C^x = J^x = \sigma^{xx} E$ and $C^t = J^t = \rho$ in the last equation gives the ohmic conductivity as
\bea \label{dilatonsigmaxx}
 \sigma^{xx} &=& \sqrt{ \left( \cT \ell^4 e^{-\phi_\star} \right)^2
 +  \left( \ell^4 e^{-\phi_\star} \right) \rho^2 /
 (g_{xx}^2)_\star } \nn\\
 &=& \sqrt{ \left( \cT \ell^4 e^{-\phi_\star} \right)^2
 + v_\star^4 \left( \ell^2 \rho \right)^2 e^{-\phi_\star} } \,,
\eea
where the last line uses the explicit form of the metric, eq.~\pref{metric}. The absence of a magnetic field and axion in this case also require vanishing Hall conductivity $\sigma^{xy} = 0$. Notice the limiting forms, depending on the relative size of $v_\star^4$ and $v_c^4 :=  e^{-\phi_\star} (\cT \ell^2/\rho)^2 \gg 1$,
%
%
\bea
 \sigma^{xx} &\simeq& \cT \ell^4 \, e^{- \phi_\star} \propto v_\star^{4} \qquad\qquad\qquad\qquad
 \hbox{if $v_\star \ll v_c$} \nn\\
 \sigma^{xx} &\simeq& v_\star^2 (\ell^2 \rho) \, e^{- \phi_\star/2} \propto v_\star^{4} \qquad\qquad\qquad\;
 \hbox{if $v_\star \gg v_c$} \,,
\eea
%
%
where the last expressions use \eqref{dilatonraddep} for a charged background. Provided $\cT \ell^4 \simeq \cO(1)$, as would be true for a D3 brane, this shows that weak coupling ({\em i.e.} $e^{-\phi_\star} \gg 1$) implies $\sigma^{xx}$ starts large --- $\sigma^{xx} \simeq \cO\left( e^{-\phi_\star} \right) \gg 1$ for $v_\star < v_c$, and then climbs to still larger values with growing $v_\star$. Notice how both regimes vary like $v_\star^4$ independent of the value of $\lambda$ for the case of the charged background.   In the case of a neutral background (where $\phi_\star$ is constant) \eqref{dilatonsigmaxx} states $\sigma^{xx}$ is independent of $v_\star$ when $v_\star \ll v_c$, but $\sigma^{xx} \propto v_\star^2$ when $v_\star \gg v_c$. As shown in Appendices C and D, for sufficiently large $\sigma^{xx}$ the probe-brane limit can eventually fail, corresponding to the need for a better approximation to understand the limit of vanishing $T$.

The temperature-dependence of this expression is encoded in the value of $v_\star$, whose determination requires a fuller specification of the metric function $h(v)$. For small $E$ we know $v_\star^{-4} \simeq v_h^{-4} \simeq C T^{4/z}$. This implies
%
%
\bea \label{dilatonsigmaxxTdep}
 \sigma^{xx} &\simeq& \frac{e^{-\phi_\star/2}}{\sqrt{C} \; T^{2/z}}
 \sqrt{ (\ell^2 \rho)^2 + C (\cT \ell^4)^2 \, e^{-\phi_\star} \, T^{4/z}} \nn\\
 &=& \frac{C'\rho}{T^{4/z}},
\eea
%
%
and so $\sigma^{xx}$ starts large for high $T$, but grows with falling temperature like $\sigma^{xx} \propto T^{-4/z}$ for very low temperatures. This shows that it is indeed small $T$ that corresponds to large $\sigma^{xx}$, and so the breakdown of the probe-brane approximation. For a charged background with $\lambda = 1$ we have $z = 5$ and so $\sigma^{xx} \propto T^{-4/5}$, while for a neutral constant dilaton background we have $\sigma^{xx} \propto T^{-2/z}$.

These two conductivies naively imply a jump in the scaling exponent in parameter space.  That is, it seems one could turn off the background charge and jump from $T^{-4/z}$ scaling to $T^{-2/z}$ scaling.  However, it is important to remember that there are two separate scaling regimes for the UV and IR in the charged background solution outlined in appendix \ref{app:chargedbackground}.  The UV solution has a constant dilaton, and so has a temperature scaling of $T^{-2/z}$ with $z=1$ similar to an uncharged background with $z=1$.  The crossover between the UV and IR depends on the size of the background charge, with the size of the IR region vanishing as the charge goes to zero.  Since the conductivity is dependent on $g_{xx}$ which is a smooth function of charge, the scaling from $T^{-4/z}$ to $T^{-2}$ is smooth as we take the background charge to zero.

\subsubsection*{Validity of the probe approximation}

It turns out that the details of the domain of validity of the probe-brane approximation differ for the cases where the background geometry describes a neutral black brane (with constant dilaton and $z$ arbitrary), or when it is that of a charged, near-extremal black brane (with a dilaton profile and an attractor behaviour). As is argued in detail in Appendix C, a necessary condition for the probe approximation is
\be \label{allancond}
 \rho \ll \left( \frac{\ell^2}{\kappa^2 L^2} \right)
 \frac{e^{- \phi_\star/2}}{v_\star^2} \,,
\ee
where $\rho$ is the charge density and $\phi_\star := \phi(v_\star)$ with $v_\star$ (defined above) approaching the horizon, $v_\star \to v_h$, for small applied electric fields, $E$. Since $v_\star \le v_h$, the probe approximation can work well right down to the horizon, $v = v_h$, provided $v_h$ is not too large (and so temperatures are not too close to zero).

For neutral branes, where $\phi$ is constant, the probe approximation ultimately fails for small enough temperatures because eventually $v_\star \simeq v_h$ is large enough to invalidate eq.~\pref{allancond}.

If, on the other hand, the source brane is charged then the above bound is more complicated because $\phi_\star$ depends nontrivially on $v_\star$ (and so also on $T$). In particular, in the very low temperature limit the near-horizon geometry can be independent of the asymptotic values of the dilaton and axion, and in the dilaton-Maxwell described above \cite{dilaton2} (and the dilaton-DBI system of Appendix D) using \eqref{dilatonraddep}, we see $e^{- \phi_\star/2} \propto v_\star^{2}$. This makes the right-hand-side of eq.~\pref{allancond} constant, and so it need not be violated at very small $T$. Appendix D explores the value, $X_h$, approached by $X$ in the near-horizon, near-extremal geometry; showing that if the background geometry is supported by a DBI action with tension $N \cT$, then $\kappa^2 N \cT/X_h > 1$, although $\kappa^2 \cT/X_h$ can be small if $N$ is sufficiently large.

\subsubsection*{Conductivities with nonzero magnetic fields}

To obtain the conductivities for general magnetic fields and asymptotic axion fields, we act on the previous result using an $SL(2,R)$ transformation. Notice in particular that this automatically ensures that the result found for $\sigma(\rho,B,T)$ has a temperature flow that commutes with the action of the group, as assumed in \S2\ to reproduce the observed phenomenology from a discrete duality group --- see Fig.~\ref{Fig:Commute}.

The transformation law, $\sigma \rightarrow (a\sigma+b)/(c\sigma+d)$, implies that the ohmic and Hall conductivities obtained starting from $\sigma^{xy}_0 = 0$ and $\sigma^{xx}_0 := \sigma_0$ (with $\sigma_0$ given in eq.~\pref{dilatonsigmaxx}) are
\be \label{xxcondtrans}
 \sigma^{xx} = \frac{\sigma_0}{d^2 + c^2 \, (\sigma_0)^2}
 \quad \hbox{and} \quad
 \sigma^{xy} = \frac{ac \, (\sigma_0)^2+bd}{d^2+c^2 \, (\sigma_0)^2} \,.
\ee
We require only the values of the parameters $a, b, c,$ and $d$ that are required to take the pure dilatonic electric case to a general axion and dyonic field.

\FIGURE[ht]{ \includegraphics[scale=0.5]{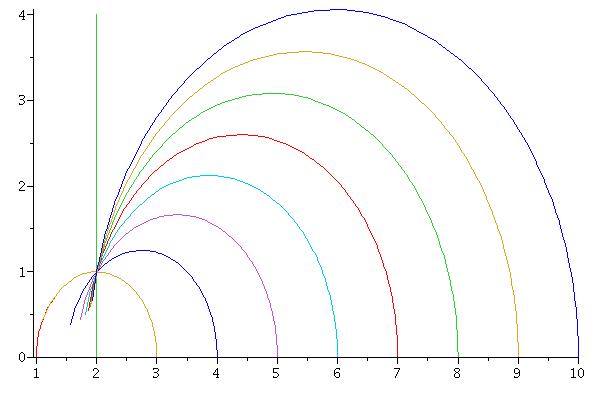}
\caption{The conductivities ($\sigma^{xx}$ plotted vs $\sigma^{xy}$), as computed using eqs.~\pref{sigmasingeneral} and \pref{sigmasingeneral2} with $\tau = 2 + i$. Each curve corresponds to a different choice for $\nu$, stepping from $\nu = 1$ to $\nu = 10$ through integer values. $\sigma_0$ is the parameter along each curve, with $\sigma^{xy} \to \nu$ in the limit of large $\sigma_0$. All lines are semi-circles centred on the real axis, and all pass through the point $\sigma = \tau$, for the reasons explained in the text (colour online).
} \label{Fig:crossedcircles} }

The required transformation is computed in Appendix A, and has parameters $a=1$, $c = -B/\rho = 1/\nu$ (where $\nu = -\rho/B$ is the filling fraction appropriate for a negatively charged particle) and
\be \label{bdformulae}
 b = \frac{\nu \left[ \chi (\nu - \chi) - e^{-2\phi} \right]}{
 (\nu - \chi)^2 + e^{-2\phi}}
 \quad \hbox{and} \quad
 d = \frac{\nu (\nu - \chi)}{(\nu - \chi)^2 + e^{-2\phi}} \,.
\ee
These lead to the conductivities
\bea \label{sigmasingeneral}
  \sigma^{xx} &=& \frac{ \nu^2 \, \left[ \left( \chi - \nu \right)^2
  + e^{-2\phi} \right]^2  \, \sigma_0 }{\nu^4
  \left( \chi - \nu \right)^2
  + \left[ \left( \chi - \nu \right)^2 + e^{-2\phi} \right]^2
  (\sigma_0)^2} \\
  \label{sigmasingeneral2}
  \sigma^{xy} &=& \frac{ \nu
  \left[ \left(\chi - \nu \right)^2
  + e^{-2\phi} \right]^2 (\sigma_0)^2 + \nu^4
  (\chi - \nu ) \left[ \chi \left( \chi - \nu \right)
  +  e^{-2\phi} \right]}{\nu^4 \left( \chi - \nu \right)^2
  +  \left[ \left(\chi - \nu
  \right)^2 + e^{-2\phi} \right]^2 (\sigma_0)^2 } \,.
\eea
where $\sigma_0$ is the $\rho$- and $T$-dependent, but $B$-independent, result given in eq.~\pref{dilatonsigmaxx} (corresponding to the $\nu \to \infty$ limit of $\sigma^{xx}$). The temperature-dependence is simplest to describe in the regime of small $E$, in which case eq.~\pref{dilatonsigmaxxTdep} can be used. In particular, for small temperatures in this case $\sigma_0 = C' \rho/T^{4/z}$ (or $\propto 1/T^{2/z}$ for neutral branes) and so is large for small $T$.

These expressions are graphed in Fig.~\ref{Fig:crossedcircles}, which plots $\sigma^{xx}$ on the vertical axis against $\sigma^{xy}$ on the horizontal. Each curve corresponds to an integer choice for $\nu$, stepping between the values $\nu = 1$ and $\nu = 10$, while the parameter $\sigma_0$ varies along each curve. Each curve approaches $\sigma^{xy} = \nu$ in the large-$\sigma_0$ limit (see below), and is a semi-circle centred on the $\sigma^{xx} = 0$ axis that passes through the point $\sigma^{xy} = \chi$ and $\sigma^{xx} = e^{-\phi}$ (so $\sigma = \tau$). Each is a semi-circle because it is the image under $SL(2,R)$ of the straight line $\sigma^{xy} = 0$, obtained for $\chi = B = 0$. Each curve passes through $\sigma = \tau$ because $\sigma$ and $\tau$ transform the same way under $SL(2,R)$ and there is always a choice for $\sigma_0$ for which the initial value of $\sigma^{xx}$ agrees with $e^{-\phi}$.

There are several limits for which the conductivities take a particularly simple form.

\vspace{4mm}\noindent
1. If $e^{-2\phi} \gg \nu^2, (\chi - \nu)^2$ (or if $\chi = \nu$, or $\nu \ll 1$, or if $\sigma_0$ is sufficiently large) then
\be \label{sigmasingeneralLargesigmaxx0}
  \sigma^{xx} = \frac{ \nu^2 }{\sigma_0}
  \quad \hbox{and} \quad
  \sigma^{xy} = \nu
 \,.
\ee
In particular, unless $\nu$ or $\chi$ are taken to be parametrically large, this result holds to the extent that we neglect loop corrections, which are controlled by powers of $e^\phi$. In particular, using the large-$\sigma_0$ limit obtained at small $T$ (for a charged background) gives the form:
\be \label{sigmasingeneralSmallT}
  \sigma^{xx} =  \frac{ \nu^2
  }{ \sigma_0  } = \frac{\rho \, T^{4/z}}{C' B^2}
  \quad \hbox{and} \quad
  \sigma^{xy} = \nu = -\frac{ \rho }{B } \,.
\ee

\vspace{4mm}\noindent
2. The limits of weak and strong magnetic field are also simple. Weak magnetic field corresponds to $\nu \to \infty$, which gives
\be \label{sigmasingeneralSmallB}
  \sigma^{xx} \to  \sigma_0 \left[ 1 - \frac{2 \chi}{\nu} + \cdots \right]
  \quad \hbox{and} \quad
  \sigma^{xy} \to \chi + \frac{(\sigma_0)^2 - e^{-2\phi}}{\nu} + \cdots  \,,
\ee
where the ellipses denote terms that are of relative order $\chi^2/\nu^2$, $e^{-2\phi}/\nu^2$ and $\sigma_0^2/\nu^2$. This generalizes the calculation of the previous section to nonzero $\chi$. By contrast, both conductivities vanish, $\sigma^{xx} = \sigma^{xy} = 0$, in the limit of large $B$ (or vanishing density) corresponding to $\nu \to 0$. The approach to zero for small $\nu$ is given by eq.~\pref{sigmasingeneralLargesigmaxx0}.

\subsection{Plateaux, semi-circles and the low-temperature limit}

Although remarkable, at face value the formulae of eqs.~\pref{sigmasingeneral} and \pref{sigmasingeneral2} do not generically describe quantum Hall plateaux, which should have vanishing ohmic conductivity, $\sigma^{xx} = 0$, combined with the defining plateau behaviour for which $\sigma_{xy}$ does not change as $B$ varies. By contrast, the generic low-temperature limit of the above formulae produce a Hall conductivity that takes a continuous range of values, $\sigma^{xy} = \nu$, as $B$ is varied, and so does not show the characteristic plateau-like feature of remaining constant as $B$ varies over a finite range. As a result, at low temperatures and magnetic fields the fluid has an ohmic conductivity that tracks the temperature and a Hall conductivity that tracks the magnetic field (or filling fraction), as shown in the left panel of Fig.~\ref{Fig:SCsigmaGrid}.

\FIGURE[ht]{ \includegraphics[scale=0.15]{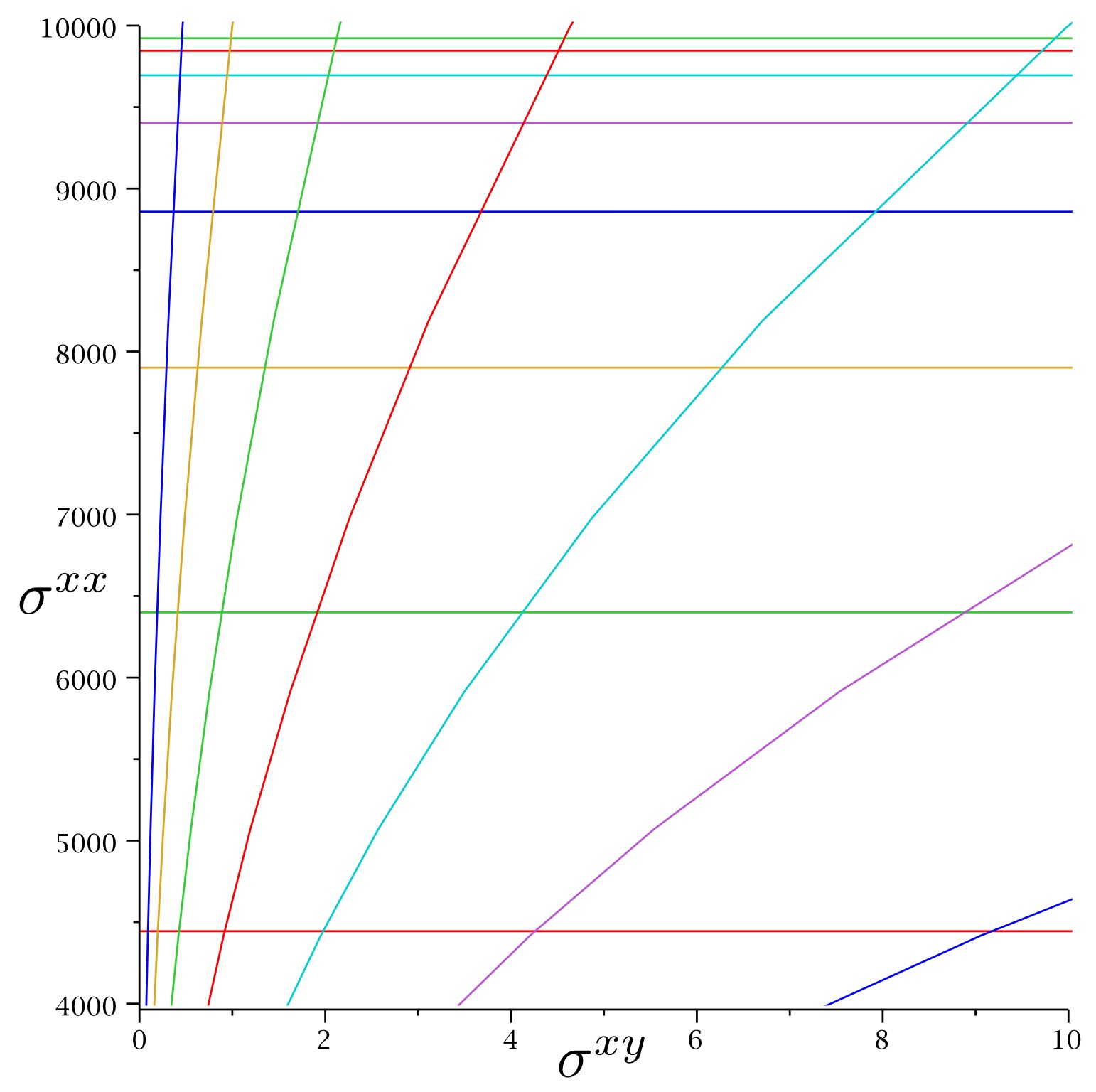}
\includegraphics[scale=0.35]{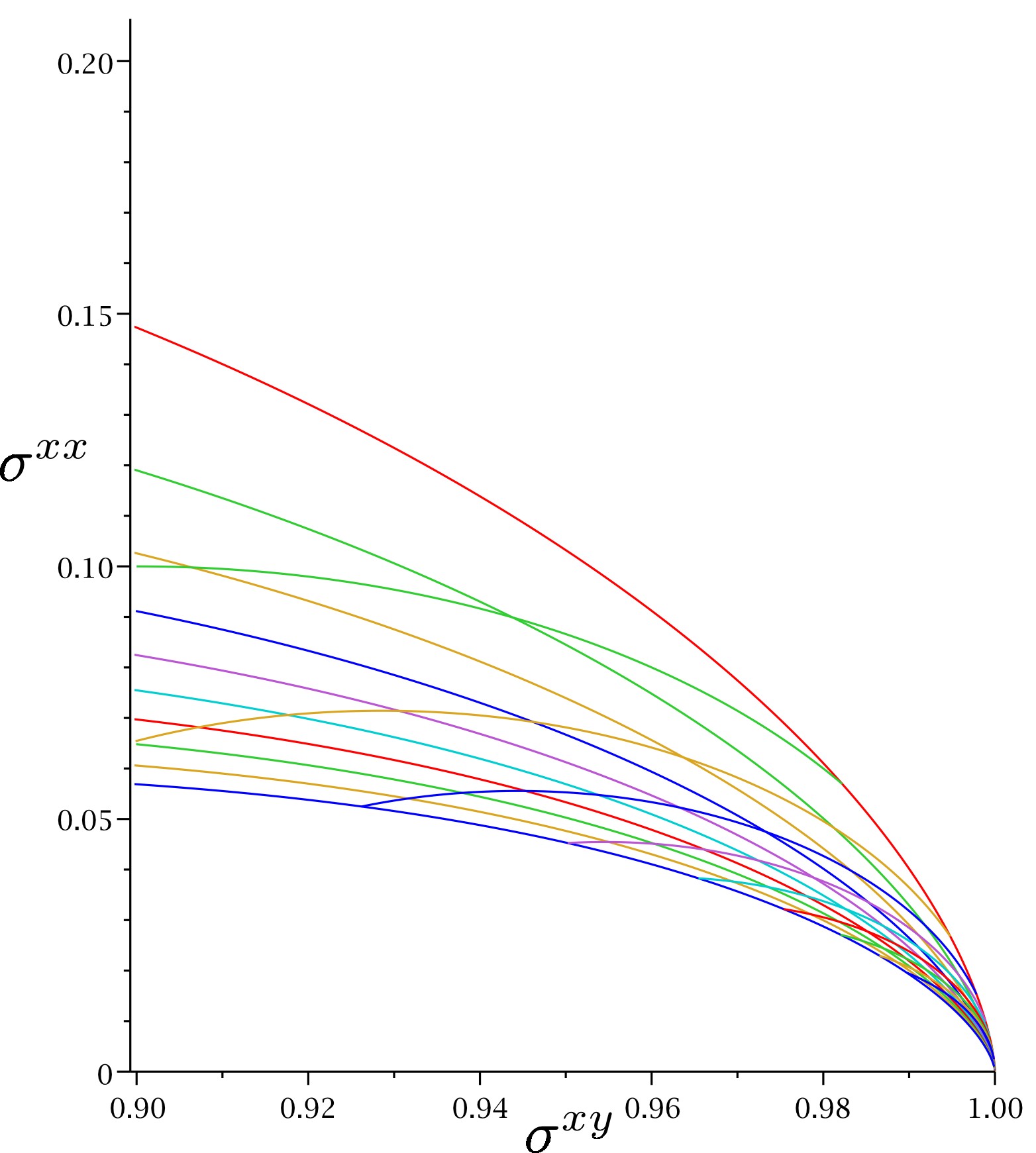}
\caption{Left panel: Curves of constant $\sigma_0$ and $\nu$ as computed semiclassically using the holographic model in the regime $\nu \gg \sigma_0^2, \, e^{-2\phi} \gg 1$. Horizontal lines represent loci of fixed $\sigma_0$ (and so also temperature), while the sloped lines describe those of fixed $\nu$ (and so also fixed magnetic field). Right panel: the same curves mapped to the strongly interacting near-plateau regime using an element of $SL(2,Z)$. The semi-circles radiating from the tip of the fan at the real axis represent lines of constant $B$, along which $T$ varies. Those transverse to these are lines of constant $T$. These illustrate a plateau behaviour inasmuch as all curves converge to the same values of $\sigma^{xx}$ and $\sigma^{xy}$ for all values of magnetic field at low temperatures (colour online).
} \label{Fig:SCsigmaGrid} }

The special point, $\sigma = \tau$, in Fig.~\ref{Fig:crossedcircles} where the many semi-circles cross is more plateau-like, however. It is plateau-like in the following specific sense: once the temperature is adjusted to sit at the point $\sigma = \tau$, changes in $\nu$ do not change the value of the conductivity. What differs between this point and those observed in quantum Hall systems is that for real systems the ohmic conductivity should also vanish, corresponding to taking Im $\tau = e^{-\phi} \to 0$. Although plateaux with Im $\tau \ne 0$ cannot describe quantum Hall systems, it would be of great interest to compute their full electromagnetic response to better understand their properties.

Clearly real quantum Hall systems (with Im $\tau \to 0$) cannot be captured by the semiclassical limit, for which $SL(2,R)$ is a good symmetry. Another hint that strong coupling should play a role comes from the recognition that the classical near-horizon configuration for $e^{-\phi}$ vanishes for extremal magnetic black holes. Similarly, ref.~\cite{dilaton2} computes the compressibility of the fluid for the dilaton-Maxwell system of \S3, and find that it is generically compressible, but would be incompressible at strong coupling if the weak-coupling formulae were simply formally extrapolated into the strong-coupling regime.

Happily, there is a way to probe strong coupling if it is assumed that a discrete symmetry like $\Gamma = PSL(2,Z)$ (or one of its subgroups) survives in the strong-coupling limit. In this case the behaviour near $\sigma^{xx} = 0$ is often the image under $\Gamma$ of a calculable region with much larger $\sigma^{xx}$ for which the above calculations are valid. This is possible to the extent that it is only the weak-coupling approximation that fails, since this is controlled by $e^{-\phi} = \hbox{Im}\, \tau \gg 1$ and $\Gamma$ maps regions with large Im $\tau$ to regions where it is small (precisely as it does for Im $\sigma$).

For instance, imagine starting from $\sigma = i \sigma_0$ at $B = 0$ and performing the transformations, eq.~\pref{xxcondtrans}, with $a = p$, $b = r$, $c = q$ and $d = s$ restricted to be integers, which yields
\be \label{xxcondtransint}
 \sigma^{xx} = \frac{\sigma_0}{s^2 + q^2 \, (\sigma_0)^2}
 \quad \hbox{and} \quad
 \sigma^{xy} = \frac{pq \, (\sigma_0)^2+rs}{s^2+q^2 \, (\sigma_0)^2} \,,
\ee
where the domain of validity is large $\sigma_0$, as before, and the assumed exact validity of the discrete transformation. In particular, although we cannot compute the explicit $T$-dependence of $\sigma_0$ very close to $T=0$, we need not be able to do so in order to explore the implications of the $SL(2,R)$ and $SL(2,Z)$ transformations so long as $\sigma_0 \to \infty$ as $T \to 0$. In this limit
\be \label{sigmanu=1}
 \sigma^{xx} \simeq \frac{1}{q^2 \sigma_0} \left[ 1 + \cO \left(
 \frac{1}{\sigma_0^{2}}
 \right) \right]
 \quad \hbox{and} \quad
 \sigma^{xy} \simeq \frac{p}{q} \left[ 1 + \cO \left( \frac{1}{\sigma_0^{2}}
 \right)  \right] \,,
\ee
where $\sigma_0 \gg 1$. These show that as $T \to 0$ the Hall conductivity, $\sigma^{xy}$, assumes a $B$ and $T$-independent quantized fractional value, $p/q$, while $\sigma^{xx}$ vanishes. The behaviour near this point as $\nu$ and $\sigma_0$ are varied over values $\nu \gg \sigma_0 \gg 1$ is illustrated on the right-hand panel of Fig.~\ref{Fig:SCsigmaGrid}, which plots the image of the left panel under the discrete transformation, $\sigma \to \sigma/(\sigma + 1)$, that maps $\sigma = \infty$ to $\sigma = 1$.

These fractional values become {\em bona fide} quantum Hall plateaux if we also take $e^{-\phi} \to \infty$ together with $\sigma_0 \to \infty$, since then the same discrete transformation that maps $\sigma \to \sigma' = \sigma /(\sigma +1)$ also takes $\tau \to \tau' = \tau/(\tau + 1)$ as well as $\nu \to \nu' = \nu/(\nu + 1)$. This ensures that the $\nu'$-independent plateau at $\sigma' = \tau'$ occurs for $\hbox{Im}\, \tau' = \hbox{Im} \, \sigma' = 0$, rather than off in the interior of the $\sigma$-plane as was the case for Fig.~\ref{Fig:crossedcircles}.

The precise values of $p$ and $q$ appearing in the fraction depend on the discrete group that is assumed to be valid, and ref.~\cite{Witten} argues this generically to be $SL(2,Z)$ (generated by $S$ and $T$ -- see Appendix A) in the presence of a spin structure, or $\Gamma_\theta(2) \subset SL(2,Z)$ (generated by $S$ and $T^2$) for no spin structure. In neither case does the above expression agree with real non-degenerate quantum Hall systems, since $SL(2,Z)$ allows arbitrary $p$, $q$, $r$ and $s$, subject only to $ps - qr = 1$ and $\Gamma_\theta(2)$ requires both $r$ and $q$ to be even (in which case $p$ and $s$ must be odd), or both $r$ and $q$ to be odd (with both $p$ and $s$ even). Both cases allow even $q$, unlike the usual situation in Zeeman-split quantum Hall systems.\footnote{There is evidence for some Hall states with even denominators, but these are the exception rather than the rule in the absence of more than one electron label (like spin, or layer number for bilayers or band label in graphene), 
for which $SL(2,Z)$ is the appropriate group.}

In particular, for spin-split systems with unbroken $\Gamma_\theta(2)$ symmetry this predicts Hall plateaux at fractions $\sigma^{xy} = p/q$ where $p$ is odd and $q$ is even, or with $p$ even and $q$ odd. This is precisely the duality group and Hall plateaux predicted \cite{PVD} for bosonic Hall systems, described in \S2\ --- {\em c.f.} eq.~\pref{bosonplateaux} --- suggesting the CFT is a strongly coupled analog of scalar electrodynamics.

\subsubsection*{Fermionic quantum Hall systems}

Given this identification of the the CFT as a bosonic Hall system, it is clear what is required to obtain a fermionic candidate to describe real quantum Hall systems. This is obtained from a bosonic system by coupling to a boundary statistics field having an odd statistics parameter, $\vartheta = \pi$, as in \S2. Within the present framework this is most easily done by performing the $SL(2,Z)$ transformation, eq.~\pref{addflux}, that implements the addition of such a flux: $g = S\, T^{-1} S$. The duality group that survives to strong couplings for the fermionic system is then $\Gamma_\ssF = g \Gamma_\ssB g^{-1}$, where $\Gamma_\ssB$ is the corresponding group in the boson system before the addition of the statistics flux. Assuming, as before, that $\Gamma_\ssB = \Gamma_\theta(2)$ for the bosonic system leads to the fermionic group $\Gamma_\ssF = g \Gamma_\theta(2) g^{-1} = \Gamma_0(2)$ (generated by $S\,T^2S$ and $T$), as is shown in Appendix A. This is precisely the group (defined by the condition that $q$ 
be even, and so for which $p$ and $s$ must also be odd) multiply proposed over the years \cite{LutkenRoss,KLZ,PVD} as providing a good phenomenological description of quantum Hall systems.

To find the $T$ and $B$ dependence of the conductivities in this case, first act on the initial bosonic conductivity with $g = S\, T^{-1} S$, for which $p = s = q = -1$ and $r = 0$. Starting with the dilaton-DBI result without a magnetic field, $\sigma^{xy}_0 = 0$ and $\sigma^{xx}_0 = \sigma_0$ given by eq.~\pref{dilatonsigmaxxTdep}, then gives the fermionic archetype:
\be \label{fermionbasic}
 \sigma^{xx}_0 = \frac{\sigma_0}{1 + \sigma_0^2}
 \quad \hbox{and} \quad
 \sigma^{xy}_0 = \frac{\sigma_0^2}{1 + \sigma_0^2} \,,
\ee
which in the low-temperature regime approaches an integer quantum Hall level, $\sigma^{xy}_0 \to 1$.

The expression for general $\chi$ and $B$ can then be found in one of two equivalent ways. One can either directly act with $S\, T^{-1} S$ on the general bosonic result, eqs.~\pref{sigmasingeneral} and \pref{sigmasingeneral2}; or one can act on eq.~\pref{fermionbasic} using the fermion $SL(2,R)$ transformation obtained by conjugating the boson duality transformation, \pref{bdformulae}, using $g = S\, T^{-1} S$. This latter is obtained by using
\be
 g \left( \begin{array}{cc}
       a & b \\
       c & d \\
     \end{array} \right) g^{-1} =
   \left( \begin{array}{ccc}
       a-b && b \\
       a-b+c-d && b+d \\
     \end{array} \right) \,.
\ee

\FIGURE[t]{ \includegraphics[scale=0.25]{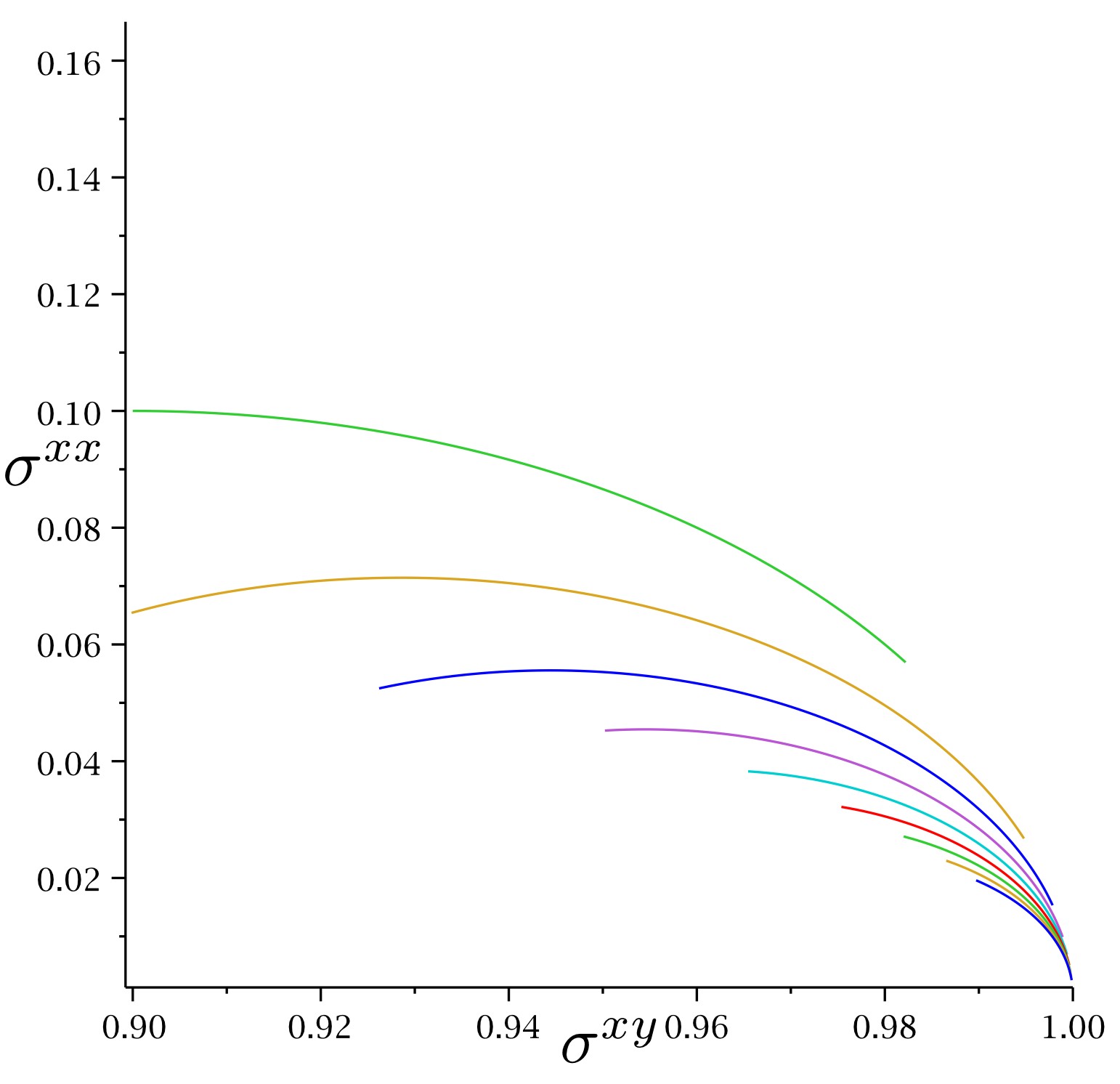}
\includegraphics[scale=0.25]{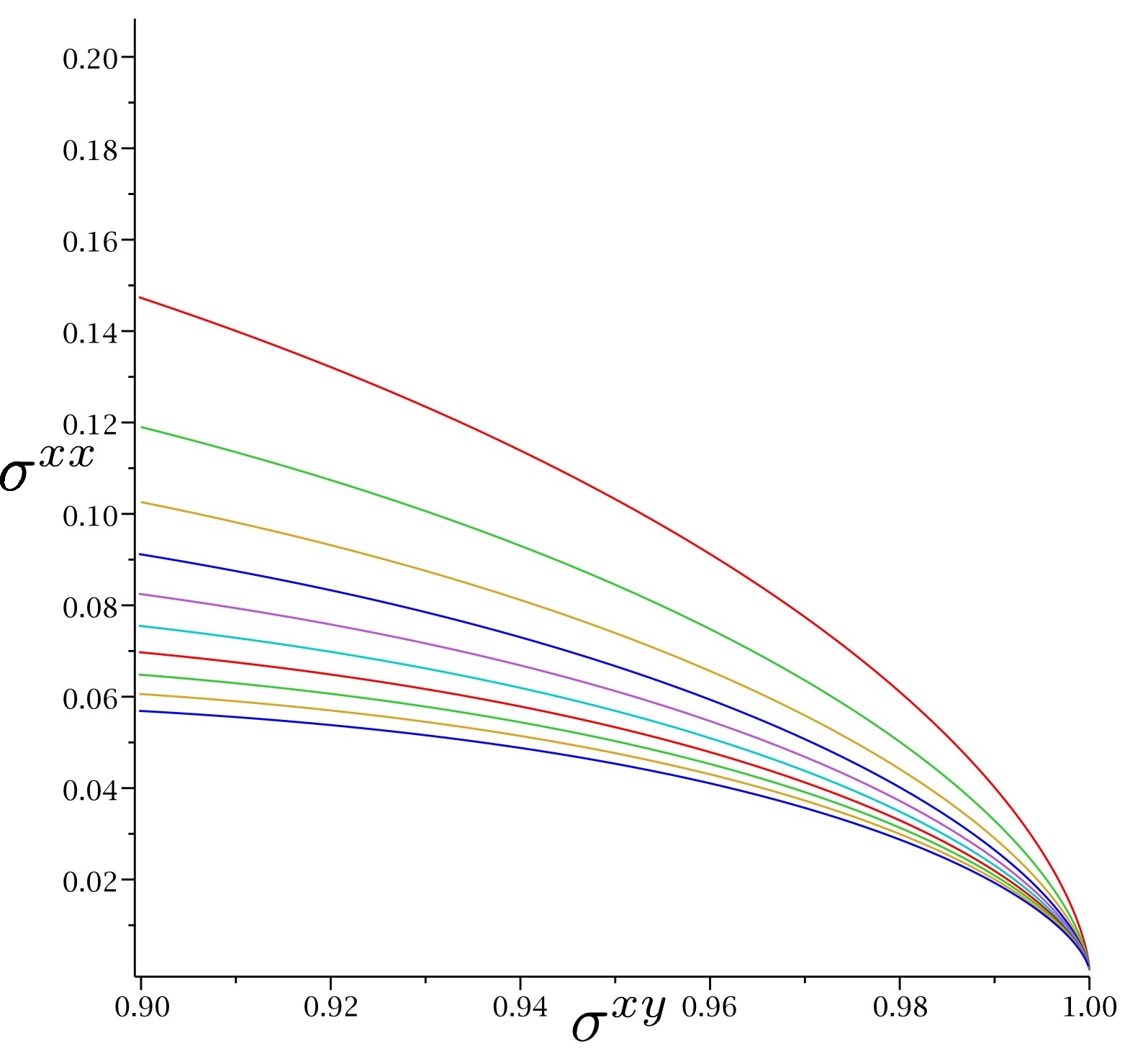}
\caption{Left panel: Curves of constant magnetic field (or $\nu' = g(\nu)$) as computed semiclassically using the holographic model and mapped onto a plateau using an element  $g \in PSL(2,Z)$. Right panel: the same curves for lines of constant $T$ (or $\sigma_0$). Constant $B$ lines are semicircles, while those along which $B$ varies become semicircles at sufficiently low temperatures (colour online).
} \label{Fig:Holocircles} }

The plateaux themselves for the fermionic system can be found by acting on the basic case, eq.~\pref{fermionbasic}, using a $\Gamma_0(2)$ transformation. Defining $\hat \sigma = \sigma^{xx}_0 + i \sigma^{xy}_0$, with components taken from \pref{fermionbasic}, the conductivity near a plateau is
\be
 \sigma = \frac{p \, \hat \sigma + r}{q \, \hat \sigma + s}\,,
\ee
where $q$ is even (and so $p$ and $s$ are odd). In the low-temperature limit, where $\sigma_0 \to \infty$ we have $\hat \sigma \to 1$ and so
\be
 \sigma \to \frac{p+r}{q+s} \,,
\ee
which clearly always has an odd denominator. The plateau-like behaviour is as illustrated on the right-hand panel of Fig.~\ref{Fig:SCsigmaGrid}.

\subsubsection*{Semicircles}

The generation of conductivities using $PSL(2,R)$, followed by mapping weak to strong coupling using a discrete symmetry also naturally ensures the observed semicircle behaviour as one approaches a quantum Hall plateau. This can be seen in Fig.~\ref{Fig:Holocircles}, which plots how lines of constant $\nu$ and $T$ approach the plateau. Lines of constant $\nu$ obtained in this way are always semi-circles because they are the images under $PSL(2,R)$ and $PSL(2,Z)$ of the straight line along which only $\sigma^{xx}$ varies when $B = 0$. Experiments varying $B$ at sufficiently small $T$ are also semicircles because these coincide with semicircular temperature flow lines. This can be seen in Fig.~\ref{Fig:Flowlines}.

\section{Discussion and conclusions}

We see that the DBI-based model examined here provides an example of a 3+1 dimensional gravitational system that has two desirable properties: $(i)$ it admits an $SL(2,R)$ duality group at the classical level; and $(ii)$ it has nonzero DC ohmic and Hall conductivities. The model is phenomenological, in that it is not part of an explicit string construction, but this is also unlikely to be relevant for low-energy purposes so long as all of the other string ingredients do not play an important role and so can be integrated out. These other stringy ingredients do play one important role, however, and that is to break the classical $SL(2,R)$ group down to a discrete subgroup. The general properties of 2+1 dimensional CFTs make this subgroup generically likely to be $SL(2,Z)$ in situations where a spin structure is relevant, or $\Gamma_\theta(2) \subset SL(2,Z)$ if a spin structure is not relevant.

These two properties are the minimal two things that would be required for a candidate description of low-energy quantum Hall systems, based on the phenomenological evidence in these systems for an emergent discrete duality symmetry (summarized in \S2). Any system with these properties automatically captures all of the implications of the discrete symmetry that survives in the strongly coupled regime, and in particular those enjoying an unbroken $\Gamma_0(2)$ duality group merit a closer inspection to see how well they capture other properties of real quantum Hall systems. This section discusses several kinds of observables of this type that are {\em not} simply consequences of duality.

\subsection{A model-building wish list}

There are two kinds of predictions made by the specific dilaton-DBI model examined here, that are typical of the kinds of comparisons that can be made that go beyond the implications of the duality groups.

\subsubsection*{Approach to zero temperature}

Although having a duality group commute with the RG flow to low temperatures predicts the properties of some of the trajectories, $\sigma(B,T)$, in the conductivity plane \cite{Semicircle}, it does not predict them all. The duality itself also does not predict the dependence on $T$ along the flow lines, although these are often reasonably well-measured ({\em c.f} eq.~\pref{phenodualform}, for example). Since specific models predict this dependence in detail, comparison with the measurements can help sort out those models that provide the best description.

\FIGURE[ht]{ \includegraphics[scale=0.8]{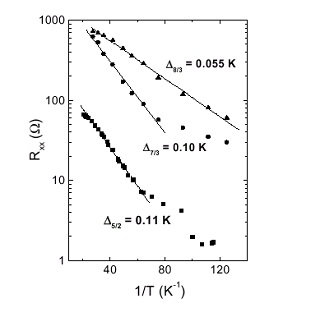}
\caption{Experimental plots, reproduced from ref.~\cite{Pan}, of the temperature dependence of the approach to various quantum Hall plateaux, showing an exponential form.
} \label{Fig:Activation} }

For instance, Fig.~\ref{Fig:Activation} shows how the ohmic DC conductivity drops exponentially with $T$ as one approaches various quantum Hall plateaux, $\propto e^{-\Delta/T}$, over a range of temperatures, crossing over to a different exponential $T$-dependence at lower temperatures, as appropriate to conductivity by hopping \cite{QHEz=1th}. Like the incompressibility of the quantum Hall state, this is consistent with the existence of a gap at low energies. By contrast, the CFT corresponding to the gravity dual described above predicts a power-law for this approach, such as the dependence $\sigma^{xx} \propto T^{4/z}$ for a charged background (or $\propto T^{2/z}$ for the uncharged background) seen near $\sigma^{xy} = 1$ in eq.~\pref{sigmanu=1}. This suggests the CFT as described so far may be limited to capturing the critical behaviour in the vicinity of a transition between two plateaux (or the transition from a plateau to the Hall insulator).

When the background describes a neutral brane, the probe-brane approximation prevents our direct exploration of the $T \to 0$ limit and so cannot exclude a crossover to exponential behaviour at very low temperatures. The same need not be true for the near-horizon extremal geometry explored in \S3\ and Appendix D, however, and does not appear to indicate such a crossover. It is possible that this exponential is associated with the approach to the low-energy 2D surface in coupling space. This motivates a more detailed study of the very low-temperature limit, as well as AdS/CFT systems having gaps like the D7 system studied in \cite{StrangeM}. We leave it as an open problem whether an alternative brane construction could be made that leads to a duality-invariant form for the low-energy 4D effective action consistent with an exponential temperature dependence.

\subsubsection*{Critical exponents}

Another experimentally accessible feature not purely dictated on symmetry grounds is the powers, $\alpha$ and $\beta$, governing the scaling of the resistivities in the low-temperature limit ({\em c.f.} eq.~\pref{rhoscaling}) and Fig.~\ref{Fig:SuperU}. As discussed in \S2, these are measured to satisfy $\beta \simeq \alpha \simeq 0.42 \pm 0.01$ \cite{Wei}, a result that can be usefully compared with the predictions of a particular CFT. How does the dilaton-DBI gravity dual described above do on this score?

The numerical equivalence $\alpha \simeq \beta$ would be easy to understand if the resistivity, $\rho_{ab}$, near the critical field, $B = B_c + \Delta B$, depended only on $T$ and $\Delta B$ through the one scaling combination
\be \label{simplescaling}
 \rho_{ab} \simeq \rho_{ab} \left( x \right)
 \quad \hbox{with} \quad
 x := \frac{\Delta B}{T^p}  \,,
\ee
for some power $p$. This dependence implies
\be
 \left( \frac{\exd \rho_{xy}}{\exd B} \right)_{B_c}
 = \frac{1}{T^{p}} \, \left( \frac{\exd \rho_{xy}}{\exd x} \right)_{x=0} \,.
\ee
Similarly, if $\Delta B$ is defined by the shape difference $\Delta \rho_{xx} = \rho_{xx}(x+ \Delta x) - \rho_{xx}(x - \Delta x)$, for a fixed $\Delta x$, then
\be
 \Delta B \propto T^p \, \Delta x \,.
\ee
Comparing these two equations gives the prediction $\alpha = \beta = p$.

Does this follow from the CFT explored above, and if so what is the predicted numerical size of $p$? In the present instance $\sigma^{ab}$ and $\rho_{ab}$ come as functions of $\nu$ and $\sigma_0$, and for small $E$ we have $\sigma_0 = \sigma_0(C'\rho/T^{4/z})$ for a charged background or $\sigma_0 = \sigma_0(\rho/T^{2/z})$ for a neutral background. For us $z$ is a parameter that is free to be dialed by adjusting $\lambda$ or the `Lifshitz sector' that sets the background geometry, but the prediction $\sigma_0 = \sigma_0(\rho/T^{2/z})$ eventually breaks down for sufficiently small $T$ for a neutral brane. Alternatively, for a charged background brane this prediction survives to lower temperatures. In either case we have the following scaling form for the conductivities
\be
 \sigma^{ab} = \sigma^{ab}( \nu, \sigma_0)
  = \sigma^{ab} \left( \frac{\rho}{T^{2/z}} \,, \frac{B}{T^{2/z}} \right) \,,
\ee
where the same power of $T$ appears with both the charge density, $\rho$, and magnetic field, $B$, because both have no anomalous dimension (since both $J^\mu$ and $\epsilon^{\mu\nu\lambda} F_{\nu\lambda}$ are conserved currents in 2+1 dimensions). Near a critical field, $B = B_c + \Delta B$, this is a function of two variables
\be
 \sigma^{ab} \simeq \sigma^{ab} \left( \Delta \nu, \frac{\Delta B}{T^{2/z}} \right) \,,
\ee
%
%
and so has the form of eq.~\pref{simplescaling} only if $\Delta \nu$ can be fixed to be small enough that $\Delta B/T^{2/z} \gg \Delta \nu$. When this is so, we would expect $\alpha \simeq \beta \simeq p \simeq 2/z$. However, it is not clear that this is the regime appropriate to the experiments that measure the power, $p$, near a transition between two plateaux, since for these $\nu$ is not fixed as $B$ varies to measure $\exd \rho_{xx}/\exd B$ near a critical transition. If this were the appropriate limit, then the observed exponent, $p \simeq 0.4$, would correspond to $z \simeq 5$ for the dynamical exponent; a value inconsistent with other measurements \cite{QHEz=1sc, QHEz=1th}.

More generally, the calculation of scaling behaviour near a critical point requires the calculation of the eigenvalues of the derivative, $\exd \beta/\exd \sigma$, near the critical conductivity, $\sigma_c$, where $\beta := T \exd \sigma/\exd T$ and $\sigma = \sigma^{xy} + i \sigma^{xx}$, say. The symmetry of the flow with respect sub-modular group ensures $\sigma_c = \frac12(1 +i)$ (or its image under the group), and so generically requires working at large coupling. Furthermore, unlike for the plateaux themselves, there are no group elements that map $\sigma_c$ out to large values of Im $\sigma$, and so to weak coupling, and so in general it is not possible to use the symmetry to compute the exponents in a weak-coupling regime. On the positive side, this means that these exponents need not be as given in mean-field theory (which would be a bad description of the experiments), but on the negative side it makes them difficult to compute explicitly.

\subsection*{Summary}

Quantum Hall systems are characterized by an impressive suite of phenomena --- quantization of the Hall conductivity; selection rules for allowed transitions between plateaux; semi-circle behaviour; $\rho_{xx} \to 1/\rho_{xx}$ duality --- that control the properties of, and the transitions between, quantum Hall plateaux. The observational evidence for these phenomena is remarkably robust; more robust than the extant theoretical explanations that are based directly on the detailed dynamics of the underlying electrons.

All of these phenomena would be robustly explained if the very low-energy approach to the quantum Hall plateaux were controlled by the RG flow through an approximately two-dimensional subspace of the space of couplings, that commutes with the duality group $\Gamma_0(2) \subset SL(2,Z)$. There is good evidence that duality groups of this type can robustly emerge within 2+1 dimensional CFTs. What has been missing is an explicit class of CFTs within which this hypothesis can be made precise, and compared in more detail with the extant experiments.

The advent of AdS/CFT models including both a discrete duality group and nonzero DC ohmic and Hall conductivities opens up the first class of models of the required type, and so opens up a new way to describe the low-energy behaviour of quantum Hall systems. We provide an explicit calculation of the DC conductivities in a simple example of this class, and describe its predictions for the low-energy approach to the quantum Hall plateaux. It generically predicts an approach for which the ohmic conductivity vanishes as a power of temperature, which is a good description of the critical behaviour, but does not capture the gapped approach to plateaux at low energy. 

The dilaton-DBI model examined here has several attractive ingredients likely to be worth incorporating into future AdS/CFT modeling of quantum Hall systems: the presence of the $SL(2,R)$ symmetry, broken by quantum effects to $SL(2,Z)$ or a subgroup; a DBI-like dynamics that naturally incorporates nonlinear effects that go beyond linear response; and the attractor near-horizon, near-extremal geometry that can make universal predictions (like for critical exponents) for very low energies if taken beyond the probe approximation. Its main drawback is the necessity to work within the probe approximation to obtain a DC resistance, requiring the invocation of a separate `Lifshitz' sector whose sole purpose is to generate the same background geometry.

But more interesting than this particular example is probably the opening up of a class of modular models, within which the drawbacks can be removed and a variety of more detailed comparisons with observations can begin to be explored.

\section*{Acknowledgements}

We thank Shamit Kachru for giving a head's up about ref.~\cite{dilaton2}, and Sean Hartnoll, Gary Horowitz, Clifford Johnson, Rob Myers, Al Shapere and Boris Shklovskii for useful conversations. CB thanks summer students A. Chan, U. Hussein, Z.Y. Niu and Y.F. Wang for their help, and the Aspen Center for Physics for providing the spectacular environment where parts of this work were done, and BD thanks McMaster University and the Perimeter Institute for hospitality as work progressed. This research has been supported in part by funds from the Natural Sciences and Engineering Research Council (NSERC) of Canada. Research at the Perimeter Institute is supported in part by the Government of Canada through NSERC and by the Province of Ontario through the Ministry of Research and Information (MRI).

\appendix

\section{Some useful properties of $SL(2,R)$ and $SL(2,Z)$}

The purpose of this appendix is to group together useful facts about the groups $SL(2,R)$, $PSL(2,R)$ and their subgroups.

The group $SL(2,R)$ consists of real-valued (or, for $SL(2,Z)$, integer-valued) two-by-two matrices with unit determinant:
\be
 M := \left( \begin{array}{cc}
     a & b \\
     c & d \\
   \end{array} \right) \,,
\ee
where det $M = 1$ requires $ad - bc = 1$.

\subsubsection*{The group $PSL(2,R)$}

Complex quantities can contain the action of this group through fractional-linear transformations,
\be \label{app:fraclin}
 z \to \frac{a \, z + b}{c \, z + d} \,.
\ee
As is easily checked, repeated applications of this transformation rule reproduces the same group multiplication law as is obtained by multiplying the matrix representation $M$. Because eq.~\pref{app:fraclin} is invariant under a simultaneous change of sign in all four parameters, $a$, $b$, $c$ and $d$, it is more properly regarded as a realization of the group $PSL(2,R)$ obtained from $SL(2,R)$ by identifying group elements that are related by $M \to - M$.

The real and imaginary parts of eq.~\pref{app:fraclin} arise often in the main text, and are given by
\bea \label{app:psl2rretrans}
 z_1 &\to& \frac{ac \, (z_1^2 + z_2^2 ) + (ad + bc) z_1 +
 bd}{c^2 ( z_1^2 + z_2^2 ) + 2cd z_1 + d^2} \\
 \label{app:psl2rimtrans}
 z_2 &\to& \frac{z_2}{c^2 ( z_1^2 + z_2^2) + 2cd \, z_1 + d^2},
\eea
where $z := z_1 + i z_2$. The second of these equations is simplified using $ad - bc = 1$.

\subsection*{The group $SL(2,Z)$ and some of its subgroups}

Any element of the group obtained when the elements of $M$ are integer-valued can be generated as a product of powers of two specific elements, traditionally called
\be
 S := \left(  \begin{array}{cc}
          0 & 1 \\
          -1 & 0 \\
        \end{array}  \right)
 \quad \hbox{and} \quad
 T := \left(  \begin{array}{cc}
          1 & 1 \\
          0 & 1 \\
        \end{array}  \right) \,,
\ee
for which the fraction-linear transformation, \pref{app:fraclin}, becomes
\be
 S(z) = - \frac{1}{z}
 \quad \hbox{and} \quad
 T(z) = z + 1 \,.
\ee
Direct matrix multiplication shows these have the property $(S \, T)^3 = 1$.

Regarded as acting on the complex variable $z$, the group $PSL(2,Z)$ maps the upper half-plane onto itself since both $S$ and $T$ preserve the sign of $z_2$. Any point in the upper half-plane can be reached from a `fundamental domain', which can be taken as the intersections of the regions $- \frac12 \le z_1 \le \frac12$ and $|z| \ge 1$.

\subsubsection*{The subgroup $\Gamma_\theta(2)$}

The subgroup\footnote{Our notation is taken from \cite{koblitz}.} $\Gamma_\theta(2)$ can be defined as that subgroup of $SL(2,Z)$ that is generated by $S$ and $T^2$, rather than $S$ and $T$. Since $T^2$ written explicitly is
\be
  T^2 := \left(  \begin{array}{cc}
          1 & 2 \\
          0 & 1 \\
        \end{array}  \right) \,,
\ee
it is clear that both $S$ and $T^2$ have the property that either $b$ and $c$ are both odd, or they are both even. Since this property is preserved under matrix multiplication, it is true for all of the elements of $\Gamma_\theta(2)$, and it can be regarded as an alternative definition of the group.

A fundamental domain for $\Gamma_\theta(2)$, from which the entire upper half-plane can be generated, can be taken as the intersection of the regions $-1 \le z_1 \le 1$ and $|z| \ge 1$.

\subsubsection*{The subgroup $\Gamma_0(2)$}

The subgroup of $SL(2,Z)$ whose properties are relevant to fermions (and so to real quantum Hall systems) is $\Gamma_0(2)$. It can be defined, as in the main text, as that group obtained by conjugating the elements of $\Gamma_\theta(2)$ by the element $g = S\, T^{-1} S \in SL(2,Z)$:
\be
 \Gamma_0(2) = g \, \Gamma_\theta(2) \, g^{-1} \,.
\ee
To see what the generators of $\Gamma_0(2)$ are it suffices to conjugate the two generators of $\Gamma_\theta(2)$, to get:
\bea
 g \, S \, g^{-1} &=& (S \, T^{-1} S) S (S \, T S) = S \, T^{-1} S \, T S
 = S \, T^{-1} (S \, T)^{-2} S \nn\\
 &=& S \, T^{-1} (T^{-1} S)^{2} S = S\, T^{-2} S \, T^{-1} \,,
\eea
which uses $(S\,T)^3 = 1$ to write $S \, T = (S \, T)^{-2}$. Similarly,
\bea
 g \, T^2 g^{-1} &=& (S \, T^{-1} S) T^2 (S \, T S) = S \, T^{-1} S \, T^2
 (S\, T)^{-2} S \nn\\
 &=& S \, T^{-1} S \, T^2 (T^{-1} S)^{2} S = S\, T^{-1} S\, T S \,T^{-1} \nn\\
 &=&  S\, T^{-2}  S\, T^{-2} \,.
\eea
But any group element that can be obtained from products of powers of these generators can equally well be generated by products of powers of the more usually chosen generators %
\be
 S \, T^2 S = \left(   \begin{array}{cc}
                  -1 & 0 \\
                  -2 & -1 \\
                \end{array}  \right)
\ee
and $T$. Notice that both $S\, T^2 S$ and $T$ have the property that the lower-left element $c$ is even, and since this is preserved under group multiplication it is true for all of the elements of $\Gamma_0(2)$. The condition $ad - bc = 1$ then implies that both $a$ and $d$ must be odd. The condition of even $c$ turns out to provide an equivalent definition of the group.

A fundamental domain for the group $\Gamma_0(2)$ can be taken as the intersections of the region $0 \le z_1 \le 1$ and $\left|z - \frac12 \right| \ge \frac12$.

\subsection*{The group element as a function of $B$, $\chi$ and $\rho$}

In \S4\ of the text the $SL(2,R)$ transformation is required that maps the special case of $B = \chi = 0$ onto the general case. This subsection determines the required transformations.

Starting with the $xy$-component of (\ref{Maxwelltransfn}), we see
\bea
  F_{xy} &=& d \, (F_{xy})_0 - c \, (\Gd_{xy})_0 \nn\\
  - \Gd_{xy} &=& b \, (F_{xy})_0 - a \, (\Gd_{xy})_0 \,,
\eea
so using $F_{xy} = B$ and $\Gd_{xy} = - \epsilon_{tvxy} G^{vt} = \sqrt{-g} \; G^{vt}= \rho$, and $B_0 = 0$, gives the relations
\begin{equation}
 c = - \frac{B}{\rho_0} \quad \hbox{and} \quad
 a = \frac{\rho}{\rho_0} \,. \label{app:arel}
\end{equation}
Finally, performing the inverse transformation to
\be \label{app:xxtautrans}
 e^{-\phi} = \frac{e^{-\phi_0}}{d^2 + c^2 \, e^{-2\phi_0}}
 \quad \hbox{and} \quad
 \chi = \frac{ac \, e^{-2\phi_0} + bd}{d^2 + c^2 \, e^{-2\phi_0}} \,.
\ee
gives
\be
 \chi_0 = 0 = \frac{-dc (\chi^2 + e^{-2\phi}) - ab + (ad+bc) \chi}{(a - c\chi)^2
 + c^2 \, e^{-2\phi}} \,,
\ee
which can be solved for $d$ once $b = (ad-1)/c$ is used, giving
\begin{equation}
 d = \rho_0 \; \left[ \frac{\rho + B \chi}{(\rho + B \chi)^2 + B^2 e^{-2\phi} }
 \right] \,,
\end{equation}
and so $b = (ad-1)/c$ is
\be
 b = \rho_0 \; \left[ \frac{\chi (\rho + B \chi ) + B \,
 e^{-2\phi}}{(\rho + B \chi )^2 + B^2 e^{-2\phi}} \right] \,.
\ee

The final form for the conductivities is therefore found by choosing $\rho = \rho_0$\footnote{As discussed in the main text, keeping $\rho \ne \rho_0$ allows the formulae to be generalized to arbitrary values for the charges of the carriers in the CFT.} and therefore $a=1$, leading to $c = -B/\rho = 1/\nu$, where $\nu = -\rho/B$ is the filling fraction (with the sign appropriate for a negatively charged particle). The remaining two parameters then are
\be
 b = \frac{\nu \left[ \chi (\nu - \chi) - e^{-2\phi} \right]}{
 (\nu - \chi)^2 + e^{-2\phi}}
 \quad \hbox{and} \quad
 d = \frac{\nu (\nu - \chi)}{(\nu - \chi)^2 + e^{-2\phi}} \,,
\ee
These are the results quoted in section \S4.

\section{DBI thermodynamics}

This section reproduces some of the thermodynamic properties of the dilaton-DBI system in the case of a neutral brane in which the dilaton is a constant considering only the brane contribution to the thermodynamics, which do not differ significantly from the non-dilaton case studied in ref.~\cite{StrangeM}.

\subsubsection*{Free energy}

The free energy (density) is found by evaluating the bulk action at the classical solution, and regarding the result as a function of the boundary values. Since the asymptotic value of $A_t$ gives the chemical potential, the result is naturally viewed as a thermodynamic potential whose variables are $T$, $\mu$, and $B$. It is convenient to instead work at fixed charge density, so we follow \cite{StrangeM} (see also \cite{HR}) by performing the Legendre transformation to obtain the potential whose natural variables are $T$, $\rho$ and $B$:
\begin{equation}
 f(T) = \frac{T S_{\rm gauge}}{V_2} + \mu J^t. \label{app:freeenergydef}
\end{equation}

For the purposes of thermodynamics it suffices to work with the following gauge field ansatz,
\begin{equation}
 A = \Phi(v) \, \exd t + B x \, \exd y.
\end{equation}
The solution to the field equations for $\Phi$ is then
\begin{equation} \label{app:densityeqn}
 F_{vt} = \Phi' = \frac{1}{v^{1+z}} \; \frac{C}{\sqrt{v^{-4} +
 \left( \frac{\ell^2}{L^2} \right)^2 (B^2+C^2)}},
\end{equation}
where $C$ is an integration constant. Eq.~\pref{app:densityeqn} can be integrated, to obtain
\begin{equation}
 \Phi(v) = \mu(v_h) + \int_\epsilon^{v} \frac{\exd \hat v}{\hat v^{1+z}} \;
 \frac{C}{\sqrt{\hat v^{-4} + \left( \frac{\ell^2}{L^2}
 \right)^2 (B^2 + C^2)}} \,.
\end{equation}
Here $\mu$ is another integration constant, to be interpreted as the chemical potential, whose value is determined by the condition that $\Phi(v_h) = 0$ at the black hole horizon. This gives the following expression for the chemical potential as a function of horizon position (temperature),
\begin{equation}
 \mu(T) = \int_\epsilon^{v_h} \frac{\exd \hat v}{\hat v^{1+z}} \;
 \frac{C}{\sqrt{\hat v^{-4} + \left( \frac{\ell^2}{L^2}
 \right)^2 (B^2 + C^2)}}. \label{app:chempot}
\end{equation}

Expanding the solution near the conformal boundary, $v=0$, gives
\begin{equation}
 \Phi = \mu - \frac{1}{v^{z-2}} \left( \frac{C}{z-2} \right) + \cdots \,,
\end{equation}
which shows that the constant, $C$ is related to the boundary charge density by
\begin{equation}
 J^t = \cT \, \ell^4 \, C \,.
\end{equation}

The free energy becomes
\begin{equation}
 f(T) = - \cT \, L^4 \int_\epsilon^{v_h} \frac{\exd v}{v^{1+z}} \;
 \frac{v^{-4} + \left( \frac{\ell^2}{L^2} \right)^2 B^2}{
 \sqrt{v^{-4} + \left(\frac{\ell^2}{L^2} \right)^2
 (B^2+C^2)}}+\mu(T) J^t \,, \label{app:fenergy}
\end{equation}
with the second term evaluated using eq.~\pref{app:chempot}. The resulting integral diverges, but since these divergences are independent of temperature they can be regulated by subtracting the free-energy at zero temperature, giving the finite result
\begin{eqnarray}
 \Delta f &:=& f(T) - f(0) \nn\\
 &=& - \cT \, L^4 \int_\infty^{v_h} \frac{\exd v}{v^{1+z}} \;
 \sqrt{v^{-4} + \left(\frac{\ell^2}{L^2} \right)^2
 (B^2+C^2)} \label{app:generalfenergydiff}\\
 &\propto& \cT \, \ell^2 \, L^2 T \sqrt{B^2+C^2} +
 \frac{\cT \, L^6 \, T^{1+4/z} }{\ell^2 \sqrt{B^2 + C^2}}
 + \cdots, \label{app:lowTfenergy}
\end{eqnarray}
where the ellipses denote higher orders in temperature. Without an exact form for $h(v)$ it is impossible to keep track of numerical factors in these expressions.

\subsubsection*{First and second order quantities}

Differentiating eq.~(\ref{app:lowTfenergy}) gives various thermodynamic quantities.  The entropy is
\begin{equation}
 S = - \frac{\partial f}{\partial T} \propto \cT \, \ell^2 \, L^2
 \sqrt{B^2 + C^2} + \frac{\cT \, L^6}{\ell^2 \sqrt{B^2+C^2}} \; T^{4/z} \,,
\end{equation}
while the specific heat is
\begin{equation}
 c_\ssV = -T \frac{\partial^2 f}{\partial T^2}
 \propto \frac{\cT \, L^6 \; T^{4/z}}{\ell^2 \sqrt{B^2 + C^2}}
  \,,
\end{equation}
at low temperatures.

The regularization described above, simply subtracting the zero-temperature result, is insufficient to render the magnetization density finite since this doesn't involve differentiating with respect to temperature. It consequently receives a contribution from the diverging zero-temperature terms. The required integral is a hypergeometric function, which at low temperatures gives
\begin{equation}
 m = - \frac{1}{V_2} \frac{\partial f}{\partial B}
 \propto \cT \, \ell^2 \, L^2 \frac{ \, T}{ \sqrt{B^2 + C^2}}
 + \cT \ell^2 L^2 B \epsilon^{-z+2} \,,
\end{equation}
where $\epsilon$ is a cutoff representing the UV sensitivity of the temperature-independent contribution. Finally the magnetic susceptibility is (at zero magnetic field)
\begin{equation}
 - \frac{1}{V_2} \frac{\partial^2 f}{\partial B^2}
 \propto \frac {\cT \ell^2 L^2 \, T}{C} +
 \cT \, \ell^2 \, L^2 \, \epsilon^{-z+2} \,,
\end{equation}
where numerical factors are not followed in the relative normalization between the two terms.

\section{Validity of the probe-brane approximation}

Here we investigate the region in which the probe brane approximation is valid, closely following \cite{StrangeM}. We vary the action with respect to $g_{tt}$ to find the energy density, and insist it must be less than the background energy density $\sim \frac{1}{\kappa^2L^2}$.

Our action (assuming ohmic conductivity and a constant electric field) takes the form
\begin{align}
 S_{\text{gauge}}&=-\mathcal T\int\exd^4x\sqrt{-g}\sqrt{1+\ell^4e^{-\phi}\left( g^{tt}g^{vv}(\Phi')^2+g^{vv}g^{xx}(\mathcal A')^2+g^{tt}g^{xx}E^2\right)}\\
 &=-\mathcal T\int\exd^4x\sqrt{-g}X.
\end{align}
Varying this action with respect to $g_{tt}$ gives
\begin{equation}
 \frac{\mathcal T}{\sqrt{-g}X}\left(g_{xx}^2g_{vv}+\ell^4 e^{-\phi}(\mathcal A')^2 g_{xx}\right),
\end{equation}
Similarly, the background energy density is found from varying $\sqrt{-g} \frac{1}{\kappa^2 L^2}$. This gives
\begin{equation}
 \frac{g_{xx}^2g_{vv}}{\sqrt{-g}}\frac{1}{\kappa^2 L^2}.
\end{equation}
The probe brane condition is therefore
\begin{equation}
 \gamma:= \frac{1}{X}\left(1+\ell^4 e^{-\phi}(\mathcal A')^2 g^{vv}g^{xx}\right)\ll \frac{1}{\kappa^2 L^2 \mathcal T}. \label{gammaconstraint}
\end{equation}
Since we're only interested in the conductivity calculation, this condition only needs to hold at $v_\star$. We evaluate $\gamma$ using equation (4.22) of the main text,
\begin{equation}
 \sqrt{-g}\frac{\mathcal T \ell^4 e^{-\phi}}{X}g^{vv}g^{xx}\mathcal A'=C^x \quad \text{and} \quad  \sqrt{-g}\frac{\mathcal T \ell^4 e^{-\phi}}{X}g^{vv}g^{tt} \Phi'=C^t,
\end{equation}
and plug this into the definition of $\gamma$.  As in the conductivity calculation, we opt to trade the functions, $\Phi$ and $\mathcal A$, for the conserved quantities, $C^t$ and $C^x$. This gives us
\begin{align}
 \gamma = & X\left(\frac{1}{X^2}-\frac{(C^x)^2 g_{xx}g_{vv}}{g\mathcal T^2 \ell^4 e^{-\phi}}\right)\\
 =& X\left(\frac{1}{X^2}-\frac{(C^x)^2 }{g_{tt}g_{xx}\mathcal T^2 \ell^4 e^{-\phi}}\right)\\
 =& \sqrt{\frac ND}\left(\frac DN -D +1+\frac{(C^t)^2}{\mathcal T^2 \ell^4 e^{-\phi}g_{xx}^2}\right) \,,
\end{align}
where
\begin{align}
 N:=&1+\ell^4 e^{-\phi}\left(\frac{E^2}{g_{tt}g_{xx}}\right)\\
 D:=&1+\frac{1}{\mathcal T^2 \ell^4 e^{-\phi}}\left[\frac{(C^t)^2}{g_{xx}^2}+\frac{(C^x)^2}{g_{xx}g_{tt}}\right]\\
 X=&\sqrt{\frac{N}{D}} \,.
\end{align}

For small electric fields the conductivity is evaluated at $v_\star \simeq v_h$. Near the horizon $g_{tt}\rightarrow 0$ and so
\begin{equation}
 X=\sqrt{\frac ND}\rightarrow \frac{\ell^4 e^{-\phi}\mathcal T}{ \sigma^{xx}} \quad \text{as }\quad v\rightarrow v_h. \label{Xhorizonlimit}
\end{equation}
This may not be true when $v_h=v_\star$ since both $N$ and $D$ vanish when $v=v_\star$, so we must be more careful in taking limits.  That is, it is not clear that the limits $v\rightarrow v_\star$ and $v\rightarrow v_h$ commute.  To find the value of $X$ at $v=v_\star$, we use l'H\^{o}pital's rule (primes denote derivatives with respect to $v$.),
\begin{equation}
 \lim_{v\rightarrow v_\star} X^2=\ell^8 e^{-2\phi}\mathcal T^2\frac{\frac{E^2}{(g_{tt}g_{xx})^2}(g'_{tt}g_{xx}+g_{tt}g'_{xx})}{2\frac{(C^t)^2}{g_{xx}^3}g'_{xx}+\frac{(C^x)^2}{g_{xx}g_{tt}}(g'_{tt}g_{xx}+g_{tt}g'_{xx})} \; \Bigg|_{v_\star}.
\end{equation}
As we take $v_\star\rightarrow v_h$, we see that $g_{tt}\big|_{v_\star}$ vanishes, while $g_{xx}\big|_{v_\star}$, $g'_{xx}\big|_{v_\star}$ and $g'_{tt}\big|_{v_\star}$ remain finite ($g'_{tt}$ remains finite since $g'_{tt}\big|_{v_h}=-h'(v_h)/v_h^{2z}\sim T/v_h^{z+1}$).  Once we use the fact that square roots commute with limits (provided the argument of the square root is positive), $X$ becomes
\begin{equation}
 \text{lim}X_{v_\star\rightarrow v_h}=\ell^4 e^{-\phi}\mathcal T \frac{E}{C^x},
\end{equation}
which is simply (\ref{Xhorizonlimit}).  Finally, we now use (\ref{Xhorizonlimit}) as well as $N=D=0$ to compute $\gamma_{v_\star}$ when $v_\star\rightarrow v_h$, which gives us
\begin{equation}
 \gamma_{v_{\star}} = \left[\left(\frac{ \sigma^{xx}}{\ell^4 e^{-\phi_\star}\mathcal T}\right)+\frac{\ell^4 e^{-\phi_\star}\mathcal T}{ \sigma^{xx}}+\frac{(C^t)^2}{\mathcal T  \sigma^{xx}g_{xx\star}^2}\right]. \label{gammacondrel}
\end{equation}
The conductivity itself is given by
\begin{equation}
 \sigma^{xx}=\sqrt{(\mathcal T \ell^4 e^{-\phi_\star})^2+\ell^4 e^{-\phi_\star}(C^t)^2/(g_{xx\star}^2)},
\end{equation}
allowing us to write $\gamma_\star$ as
\begin{equation}
 \gamma_\star=\frac{2\sigma^{xx}}{\ell^4e^{-\phi_\star}\mathcal T}.
\end{equation}
We can now put a constraint on the conductivity, $\sigma^{xx}$, instead of $\gamma_\star$.  Using (\ref{gammaconstraint}), our condition is (dropping factors of order unity)
\begin{equation}
 \sigma^{xx}\ll \frac{\ell^4e^{-\phi_\star}}{L^2\kappa^2}.
\end{equation}
To be clear, this condition was obtained by taking the limit $v\rightarrow v_\star$ to obtain the expression of the conductivity, then taking the limit $v_\star\rightarrow v_h$ for the conductivity at low temperatures. Notice how this condition on $\sigma^{xx}$ is independent of brane tension and just on the ratio of the coupling strengths of the gauge and gravity sector.

We can use this relation to put a constraint on the charge density instead.  Since we know at low temperatures, $\sigma^{xx}$ behaves as
\begin{equation}
 \sigma^{xx}\simeq v_\star^2 \ell^2\rho e^{-\phi_\star/2},
\end{equation}
there is a condition on rho,
\begin{equation}
 \rho \ll \frac{\ell^2e^{-\phi_\star/2}}{v_\star^2\kappa^2 L^2}.
\end{equation}
This ensures that we can not go to zero temperature ($v_\star\sim v_h\rightarrow \infty$) for finite values of $\rho$ without making our gauge coupling, ${e^{\phi}}/{\ell^4}$, vanish.

\subsection*{Dilaton field equations}

We can now ask what happens to the dilaton equations of motion near the horizon.  We again focus on $\chi=0$ and an ohmic conductivity with a constant electric field.  Looking at the source term of the dilaton equation,
\begin{equation}
\Box \phi=- \frac{\kappa^2\mathcal T \ell^4e^{-\phi}}{4X}\left[g^{tt}g^{vv}(\Phi')^2+g^{vv}g^{xx}(\mathcal A')^2+g^{tt}g^{xx}E^2\right],
\end{equation}
we can again trade out the gauge functions for the conserved quantities.  This makes the source of the dilaton equation,
\begin{equation}
 \Box \phi=-\frac{\kappa^2\mathcal T (X^2-1)}{4X}.
\end{equation}
We take the near-horizon limit and use (\ref{Xhorizonlimit}) to express this in terms of the ohmic conductivity,
\begin{equation}
 \frac{4\Box \phi}{\kappa^2}= -\frac{\mathcal T^2 \ell^4 e^{-\phi}}{\sigma^{xx}}+\frac{\sigma^{xx}}{e^{-\phi}\ell^4}.
\end{equation}
We see that the dilaton is driven to strong or weak coupling depending on the initial value of the dilaton and the conductivity.  With too large a conductivity, the dilaton is driven to strong coupling, ruining our probe brane approximation near the horizon.  Of course, this can always be remedied by choosing an appropriate initial value of the dilaton when integrating the equations of motion from the boudary.

\section{DBI near-horizon extremal geometry} \label{app:chargedbackground}

In this section we compute the attractor exponent $z$ for the near-horizon geometry of the extremal black hole using the dilaton-DBI action beyond the probe-brane approximation, verifying that $z = 5$ as for the dilaton-Maxwell case. Although our real interest is the near-extremal case in order to maintain calculational control, the simpler extremal geometry suffices for the purpose of identifying $z$. For simplicity we work with the purely electric black brane in dilaton-DBI gravity, with both the magnetic and axion fields set to zero. We return to generalizing to near-extremal and nonzero magnetic and axion fields at the end.

\subsection*{Action and field equations}

With the axion set to zero the action becomes
\be
 S = - \int \exd^4x \, \sqrt{-g} \; \left\{ \frac{1}{2 \kappa^2}
 \left[ R - 2 \Lambda + \frac{\lambda^2}2 \, \partial_\mu \phi \,
 \partial^\mu \phi \right] + \cT \Bigl( X - 1 \Bigr) \right\} \,,
\ee
with
\be
 X = \sqrt{ 1 + \frac{\ell^4}{2} \, e^{-\phi} F^2
 - \frac{\ell^8}{16} \, e^{-2\phi} (F \Fd)^2 } \,,
\ee
as before. The field equations for this action are
\be \label{app:maxwell}
 \nabla_\mu G^{\mu\nu} = 0  \,,
\ee
\be \label{app:dilaton}
 \lambda^2 \Box \phi + \frac{\kappa^2}{2} \, G^{\mu\nu} F_{\mu\nu} = 0 \,,
\ee
with
\be \label{app:gdef}
 G^{\mu\nu} = \frac{\cT \ell^4}{X} \left[ e^{-\phi} F^{\mu\nu}
 - \frac{\ell^4}{4} \, e^{-2\phi} (F \Fd) \Fd^{\mu\nu} \right] \,,
\ee
and
\be \label{app:einstein}
 R_{\mu\nu} + \frac{\lambda^2}2 \, \partial_\mu \phi \, \partial_\nu \phi
 - \Lambda \, g_{\mu\nu} + \frac{\kappa^2 \cT}{X} \Bigl[ \ell^4
 \, e^{-\phi} F_{\mu\lambda} {F_\nu}^\lambda - (X - 1) g_{\mu\nu}
 \Bigr] = 0 \,.
\ee

\subsection*{Radial ansatz}

We seek solutions to these equations subject to the ansatz $\phi = \phi(r)$ and $F = F_{rt} (r) \; \exd r \wedge \exd t$, with metric
\be \label{app:metric}
 \exd s^2 = - h(r) \, e^{-\xi(r)} \, \exd t^2 + \frac{\exd r^2}{h(r)}
 + r^2 \Bigl( \exd x^2 + \exd y^2 \Bigr) \,.
\ee
These coordinates are related to those in the main text by $r = 1/v$, so conformal infinity is at $r \to \infty$ and the horizon is at $r = 0$ (for an extremal black brane).

The Maxwell equation, eq.~\pref{app:maxwell}, integrates to give
\be
 G^{rt} = -\, \frac{Q_e \, e^{\xi/2}}{r^2} \,,
\ee
where $Q_e$ is an integration constant. Combining this with the constitutive relation, eq.~\pref{app:gdef}, then gives (after some algebra)
\be \label{app:maxwellsoln}
 - e^{-\phi} F^2 =  \frac{2 Q_e^2}{ Q_e^2 \ell^4 +  \left(\mathcal T \ell^4 \right)^2
 \, r^4 \, e^{-\phi}} \,.
\ee
The dilaton equation, eq.~\pref{app:dilaton}, evaluated using the above ansatz becomes
\be \label{app:dilatonr}
  \Bigl( r^2 e^{-\xi/2} \, h \, \phi' \Bigr)'
  - \frac{\kappa^2Q_e}{4} \, F_{rt}  = 0\,,
\ee
where primes denote differentiation with respect to $r$. Evaluation of the Einstein equations requires the following components of the Ricci tensor,
\bea
 g^{tt} R_{tt} &=& \frac{h''}{2} - \frac{3 h' \, \xi'}{4} - \frac{h \, \xi''}{2}
 + \frac{h \, (\xi')^2}{4} + \frac{h'}{r} - \frac{h \, \xi'}{r} \,, \nn\\
 g^{rr} R_{rr} &=& \frac{h''}{2} - \frac{3 h' \, \xi'}{4} - \frac{h \, \xi''}{2}
 + \frac{h \, (\xi')^2}{4} + \frac{h'}{r}  \,, \nn\\
 g^{xx} R_{xx} = g^{yy} R_{yy}
 &=& \frac{h'}{r} - \frac{h \, \xi'}{2\,r} + \frac{h}{r^2} \,.
\eea
With these, the $(tt) - (rr)$ Einstein equation becomes
\be \label{app:einsteintr}
 \xi' + \frac {\lambda^2}2 \, \left( \phi' \right)^2 = 0 \,,
\ee
while the $(xx)$ and $(yy)$ Einstein equations give
\be \label{app:einsteinxx}
 \frac{h'}{r} - \frac{h \, \xi'}{2\,r} + \frac{h}{r^2}
 = \Lambda + \kappa^2 \cT \left( \frac{X - 1}{X} \right) \,.
\ee

\subsection*{Solutions}

Eq.~\pref{app:einsteinxx} has a simple power-law solution for any region where $X$ is approximately constant. These resemble the known solutions \cite{Taylor} for the pure dilaton-Maxwell case, which are included here as the special case $r \to \infty$ since $e^{-\phi} \, F^2 \to 0$ implies $X \to 1$ in this regime. The power-law solution is
\be
 h \propto r^2 \quad \hbox{and} \quad
 e^{- \xi} \propto r^{\omega_\xi} \,,
\ee
for any $\omega_\xi$. Eq.~\pref{app:einsteintr} then implies $\phi$ is also described by a power law,
\be
 e^\phi \propto r^{\omega_\phi} \,,
\ee
with $\omega_\xi = \frac{\lambda^2}2 \, \omega_\phi^2$. There are two ways that $F_{rt}$ can then scale consistent with having $X$ constant, and these define the regimes of large and small $r$ that are of particular interest.

\subsubsection*{Large-$r$ regime}

The large-$r$ regime exploits the above solution by choosing $\omega_\phi = \omega_\xi = 0$, which implies $r^4 e^{-\phi} \to \infty$ as $r \to \infty$ and so --- from eq.~\pref{app:maxwellsoln} --- we have $e^{-\phi} F^2 \propto 1/r^{4} \to 0$ and so $X \to 1$. In this limit eq.~\pref{app:gdef} implies the Maxwell field falls off as
\be
 F_{rt} \propto \frac{1}{r^2} \,,
\ee
which, when used in the dilaton equation, eq.~\pref{app:dilatonr}, gives the sub-dominant fall-off: $\phi - \phi_\infty \propto 1/r^4$. This gives the asymptotic geometry
\be
 \exd s^2 \simeq h_\infty \, r^2 \, \exd t^2 + \frac{\exd r^2}{h_\infty \, r^2} + r^2 \Bigl(
 \exd x^2 + \exd y^2 \Bigr) \,,
\ee
and so defining the anisotropic exponent $z$ by $g_{tt} \propto r^{2z}$ when $g_{xx} = g_{yy} \propto r^2$ shows that $z = 1$ in the UV (large-$r$) limit.

\subsubsection*{Small-$r$ regime}

The other regime of constant $X$ is the near-horizon limit, $r \to 0$. In this case $X$ can be constant if $e^\phi \propto r^4$ as $r \to 0$, so that $r^4 e^{-\phi}$ in the denominator of eq.~\pref{app:maxwellsoln} approaches a constant. The same arguments as given above using the Einstein equations then show $h \propto r^2$ and $e^{-\xi} \propto r^{8\lambda^2} = r^{2(z-1)}$ when defining the anisotropic exponent $z$ by $g_{tt} \propto r^{2z}$. The constitutive equation, \pref{app:gdef}, then implies $F_{rt} \propto r^{z+1}$, which when used in the dilaton equation, eq.~\pref{app:dilatonr}, gives the consistency condition
\be
 \left(r^4 e^{-\xi/2} \phi' \right)' \propto r^{z+1} \,,
\ee
which is consistent with the above choices: $\phi' \propto 1/r$ and $e^{-\xi/2} \propto r^{z-1}$.

This gives the asymptotic near-horizon geometry as
\be
 \exd s^2 \simeq h_0 \, r^{10} \, \exd t^2 + \frac{\exd r^2}{h_0 \, r^2} + r^2 \Bigl(
 \exd x^2 + \exd y^2 \Bigr) \,.
\ee
Once these powers are chosen the values of the pre-multiplying constants are also fixed by the field equations. That is, if $h \simeq h_h \, r^2$, $e^\phi \simeq e^{\phi_h} \, r^4$ and $e^{-\xi} \simeq e^{-\xi_h} \, r^{2(z-1)}$ then the constants $h_h$, $\phi_h$ and $\xi_h$ can be computed using the field equations. This implies a prediction, in particular for the value, $X_h$, obtained by the function $X$ as $r \to 0$, given by
\be
  X_h = \frac{-\left(\frac{z-1}2\right)(3+\mathcal{\hat T}) + \sqrt{\left(\frac{z-1}2\right)^2(3+\mathcal{\hat T})^2 + z\mathcal{\hat T}^2}}{ \mathcal{\hat T}} \,,
\ee
where $\hat \cT := \kappa^2 L^2 \cT$ is a dimensionless brane tension. Notice this has the property that $0 \le X_h \le 1$, with $X_h$ varying from zero to unity as $\hat \cT$ varies from zero to infinity, and always satisfies $\kappa^2 L^2 \cT/X_h > 1$.

\subsection*{Generalizations}

We next record in passing several easy generalizations to these solutions.

\subsubsection*{Near-extremal black hole}

There is a simple generalization of the $r \to 0$ solution to include a nonzero temperature. This comes from the recognition that $h \propto r^2$ is not the only solution to eq.~\pref{app:einsteinxx} in the regime where $X$ is constant. Since this equation is linear in $h$ a more general solution is obtained by adding to this the solution to the homogeneous equation,
\be
 \frac{h'}{r} - \frac{h \, \xi'}{2\,r} + \frac{h}{r^2}
 = \frac{h'}{r} + \frac{zh}{r^2} = 0 \,,
\ee
where the first equality uses the same solution for $\xi$ as before: $\xi' = -2(z-1)/r$. The more general solution is then clearly
\be
 h = h_0 \, r^2 \left[ 1 - \left( \frac{r_h}{r} \right)^{z+2} \right] \,,
\ee
where the integration constant, $r_h$, denotes the nonzero position of the horizon of the now non-extremal black hole.

Since only $h(r)$ is modified, eqs.~\pref{app:einsteintr} and \pref{app:maxwellsoln} remain solved using the previous solutions $e^\phi \propto r^4$ and $e^{-\xi} \propto r^8$. Furthermore, since $F^2$ is independent of $h(r)$ it is still true that $X$ is constant for this new solution. All that remains is to check the dilaton equation, eq.~\pref{app:dilatonr}, which is easily seen to be solved because $r_h$ drops out of
\be
 \left( r^2 e^{-\xi/2} \, h \, \phi' \right)' = 4 h_0 \, e^{-\xi_0/2}
 \left\{ r^{z+2} \left[  1 - \left( \frac{r_h}{r} \right)^{z+2} \right] \right\}'
 = 28 h_0 \, e^{-\xi_0} r^{z+1} \,.
\ee

This solution provides the near-horizon, near-extremal geometry that governs the low-temperature limit. In particular, it verifies that the presence of a nonzero temperature does not change the value found earlier for $z$ in the far IR.

\subsubsection*{Nonzero axion and magnetic field}

Another trivial generalization is to act on the above near-horizon solutions with $SL(2,R)$ to generate their analogs having nonzero axion and magnetic fields. Because the field equations and Bianchi identities demand $\Gd_{xy} = \sqrt{-g} \; G^{rt} = -Q_e$ and $F_{xy} = Q_m$ are constants, it is useful to use the $SL(2,R)$ transformations
\be
 (- \Gd_{xy}) = a (- \Gd_{xy})_0 + b ( F_{xy} )_0
 \quad \hbox{and} \quad
 (F_{xy}) = c (- \Gd_{xy})_0 + d ( F_{xy} )_0 \,,
\ee
to learn $a = 1$ (if we demand we do not change $Q_e$) and $c = Q_m/Q_e$ (if we start from zero magnetic charge, $Q_m = 0$). Then after transforming the axion and dilaton become
\be
 e^\phi = c^2 \, e^{-\phi_0} + d^2 \, e^{\phi_0}
 \quad \hbox{and} \quad
 \chi = \frac{ac + bd \, e^{2\phi_0}}{c^2 + d^2 \, e^{2\phi_0}}
 \simeq \frac{Q_e}{Q_m} \left[ 1 + \cO \left( r^8 \right) \right] \,,
\ee
which uses $e^{\phi_0(r)} \propto r^4$ as $r \to 0$. This shows that $e^\phi$ is driven to strong coupling (for which the above classical arguments break down) as soon as $Q_m \ne 0$. The axion is similarly driven to $\chi \to Q_e/Q_m$ as $r \to 0$, and although this classical conclusion cannot be trusted in the strong-coupling limit, the attraction to quantized fractions is an exact consequence of unbroken $PSL(2,Z)$ (or one of its subgroups).

\end{document}